\documentclass[12pt]{article}

\usepackage{mathrsfs}
\usepackage[T1]{fontenc}
\usepackage{mathpazo}
\usepackage{setspace}
\usepackage{amsfonts}
\usepackage{amssymb}
\usepackage{amsmath}
\usepackage{esint}                   
\usepackage{epsfig}
\usepackage{latexsym}
\usepackage{color}
\usepackage{graphicx}
\usepackage{hyperref}
\usepackage{nicefrac}
\usepackage[latin1]{inputenc}
\usepackage{pstricks}
\usepackage{slashed}
\usepackage{multirow}
\usepackage{ulem}
\usepackage{comment}
\usepackage[nosort]{cite}
\usepackage{datetime}
\usepackage{tikz-cd}
\usepackage{float}

\usepackage{relsize}

\usepackage{tikz}
\usepackage{capt-of}
\usetikzlibrary{decorations.markings}

\usepackage[greek,english]{babel}
\languageattribute{greek}{polutoniko}

\usepackage[vcentermath]{youngtab}

\def\hybrid{\topmargin -30pt    \oddsidemargin 0pt 
        \headheight 0pt \headsep 0pt
        \textwidth 6.25in       
        \textheight 9.5in       
        \marginparwidth .875in
        \parskip 5pt plus 1pt   \jot = 1.5ex}

\hybrid

\def\baselinestretch{1.2}

\catcode`\@=11

\def\marginnote#1{}
\newcount\hour
\newcount\minute
\newtoks\amorpm
\hour=\time\divide\hour by60
\minute=\time{\multiply\hour by60 \global\advance\minute by-\hour}
\edef\standardtime{{\ifnum\hour<12 \global\amorpm={am}%
        \else\global\amorpm={pm}\advance\hour by-12 \fi
        \ifnum\hour=0 \hour=12 \fi
        \number\hour:\ifnum\minute<10 0\fi\number\minute\the\amorpm}}
\edef\militarytime{\number\hour:\ifnum\minute<10 0\fi\number\minute}

\def\draftlabel#1{{\@bsphack\if@filesw {\let\thepage\relax
   \xdef\@gtempa{\write\@auxout{\string
      \newlabel{#1}{{\@currentlabel}{\thepage}}}}}\@gtempa
   \if@nobreak \ifvmode\nobreak\fi\fi\fi\@esphack}
        \gdef\@eqnlabel{#1}}
\def\@eqnlabel{}
\def\@vacuum{}
\def\draftmarginnote#1{\marginpar{\raggedright\scriptsize\tt#1}}

\def\draft{\oddsidemargin -.5truein
        \def\@oddfoot{\sl preliminary draft \hfil
        \rm\thepage\hfil\sl\today\quad\militarytime}
        \let\@evenfoot\@oddfoot \overfullrule 3pt
        \let\label=\draftlabel
        \let\marginnote=\draftmarginnote
   \def\@eqnnum{(\theequation)\rlap{\kern\marginparsep\tt\@eqnlabel}%
\global\let\@eqnlabel\@vacuum}  }

\def\draft2{
        \def\@oddfoot{\sl preliminary draft \hfil
        \rm\thepage\hfil\sl\today\quad\militarytime}
        \let\@evenfoot\@oddfoot \overfullrule 3pt
        \let\marginnote=\draftmarginnote
   \def\@eqnnum{(\theequation)\rlap{\kern\marginparsep\tt\@eqnlabel}%
\global\let\@eqnlabel\@vacuum}  }

\def\preprint{\twocolumn\sloppy\flushbottom\parindent 2em
        \leftmargini 2em\leftmarginv .5em\leftmarginvi .5em
        \oddsidemargin -.5in    \evensidemargin -.5in
        \columnsep .4in \footheight 0pt
        \textwidth 10.in        \topmargin  -.4in
        \headheight 12pt \topskip .4in
        \textheight 6.9in \footskip 0pt
        \def\@oddhead{\thepage\hfil\addtocounter{page}{1}\thepage}
        \let\@evenhead\@oddhead \def\@oddfoot{} \def\@evenfoot{} }

\def\numberbysection{\@addtoreset{equation}{section}
        \def\theequation{\thesection.\arabic{equation}}}

\def\underline#1{\relax\ifmmode\@@underline#1\else
        $\@@underline{\hbox{#1}}$\relax\fi}

\def\titlepage{\@restonecolfalse\if@twocolumn\@restonecoltrue\onecolumn
     \else \newpage \fi \thispagestyle{empty}\c@page\z@
        \def\thefootnote{\fnsymbol{footnote}} }

\def\endtitlepage{\if@restonecol\twocolumn \else \newpage \fi
        \def\thefootnote{\arabic{footnote}}
        \setcounter{footnote}{0}}  

\catcode`@=12
\relax

\def\figcap{\section*{Figure Captions\markboth
        {FIGURECAPTIONS}{FIGURECAPTIONS}}\list
        {Figure \arabic{enumi}:\hfill}{\settowidth\labelwidth{Figure
999:}
        \leftmargin\labelwidth
        \advance\leftmargin\labelsep\usecounter{enumi}}}
 \relax
\def\tablecap{\section*{Table Captions\markboth
        {TABLECAPTIONS}{TABLECAPTIONS}}\list
        {Table \arabic{enumi}:\hfill}{\settowidth\labelwidth{Table
999:}
        \leftmargin\labelwidth
        \advance\leftmargin\labelsep\usecounter{enumi}}}
 \relax
\def\reflist{\section*{References\markboth
        {REFLIST}{REFLIST}}\list
        {[\arabic{enumi}]\hfill}{\settowidth\labelwidth{[999]}
        \leftmargin\labelwidth
        \advance\leftmargin\labelsep\usecounter{enumi}}}
 \relax
\makeatletter
\newcounter{pubctr}
\def\publist{\@ifnextchar[{\@publist}{\@@publist}}
\def\@publist[#1]{\list
        {[\arabic{pubctr}]\hfill}{\settowidth\labelwidth{[999]}
        \leftmargin\labelwidth
        \advance\leftmargin\labelsep
        \@nmbrlisttrue\def\@listctr{pubctr}
        \setcounter{pubctr}{#1}\addtocounter{pubctr}{-1}}}
\def\@@publist{\list
        {[\arabic{pubctr}]\hfill}{\settowidth\labelwidth{[999]}
        \leftmargin\labelwidth
        \advance\leftmargin\labelsep
        \@nmbrlisttrue\def\@listctr{pubctr}}}
 \relax
\makeatother

\def\be{\begin{equation}}
\def\ee{\end{equation}}
\def\ba{\begin{eqnarray}}
\def\ea{\end{eqnarray}}

\def\del{\partial}

\def\XXint#1#2#3{{\setbox0=\hbox{$#1{#2#3}{\int}$}
     \vcenter{\hbox{$#2#3$}}\kern-.5\wd0}}

\def\Xint#1{\mathchoice
   {\XXint\displaystyle\textstyle{#1}}%
   {\XXint\textstyle\scriptstyle{#1}}%
   {\XXint\scriptstyle\scriptscriptstyle{#1}}%
   {\XXint\scriptscriptstyle\scriptscriptstyle{#1}}%
   \!\int}

\def\r{\rho}

\def\g{\gamma}

\def\d{\delta}

\def\e{\epsilon}

\def\l{\lambda}

\def\bk{{\bf k}}
\def\bz{{\bf z}}

\def\no{\noindent}

\def\qq{\qquad}

\def\IR{\relax{\rm I\kern-.18em R}}

\def\inv{^{\raise.0ex\hbox{${\scriptscriptstyle -}$}\kern-.05em 1}}

\def \ha {{\frac{1}{2}}}

\def \ov {\over}

\def\diag{{\rm diag}}

\def\half{{\textstyle {1 \over 2}}}

\newcommand{\bb}{\hskip -0.1cm}

\newcommand{\hm}{\hskip -0.05cm - \hskip -0.05cm}

\def\tr{\textrm{Tr}}

\def\red{\textcolor[rgb]{0.98,0.00,0.00}}

\begin{document}


\renewcommand{\theequation}{\thesection.\arabic{equation}}
\csname @addtoreset\endcsname{equation}{section}

\begin{titlepage}
\begin{center}

\renewcommand*{\thefootnote}{\arabic{footnote}}

\phantom{xx}
\vskip 0.5in

{\large {\bf Phase transitions in the decomposition of $SU(N)$ representations}}


\vskip 0.4in

{\bf Alexios P. Polychronakos$^{1,2}$}\hskip .15cm and \hskip .15cm
{\bf Konstantinos Sfetsos}$^{3}$

\vskip 0.14in

${}^1\!$ Physics Department, the City College of New York\\
160 Convent Avenue, New York, NY 10031, USA\\
{\footnotesize{\tt apolychronakos@ccny.cuny.edu}}\\
\vskip 0.3cm
${}^2\!$ The Graduate School and University Center, City University of New York\\
365 Fifth Avenue, New York, NY 10016, USA\\
{\footnotesize{\tt apolychronakos@gc.cuny.edu}}

\vskip .14in

${}^3\!$
Department of Nuclear and Particle Physics, \\
Faculty of Physics, National and Kapodistrian University of Athens, \\
Athens 15784, Greece\\
{\footnotesize{ ksfetsos@phys.uoa.gr}}\\

\vskip .3in
\today

\vskip .2in

\end{center}

\vskip .2in

\centerline{\bf Abstract}

\no

\vskip .2in

\no
We study the multiplicity of irreducible representations in the decomposition of $n$ fundamentals
of $SU(N)$ weighted by a power of their dimension in the large $n$ and large $N$ double scaling limit.
A nontrivial scaling is obtained by keeping $n/N^2$ fixed, which plays the role of an order parameter.
We find that the system generically undergoes a fourth order phase transition in this parameter, from
a dense phase to a dilute phase. The transition is enhanced to third order for the
unweighted multiplicity, and disappears altogether when weighting with the first power of the dimension.
This corresponds to the infinite temperature partition function of non-Abelian
ferromagnets, and the results should be relevant to the thermodynamic limit of such
ferromagnets at high temperatures.

\vfill

\end{titlepage}
\vfill
\eject



\def\baselinestretch{1.2}
\baselineskip 20 pt

\newcommand{\eqn}[1]{(\ref{#1})}

\tableofcontents


\section{Introduction}

Nonabelian unitary groups $SU(N)$ play a crucial role in particle physics \cite{Hoo,BIPZ}, 
and, indirectly through matrix models \cite{GrMi}, in string theory and gravity. Ungauged and gauged $SU(2)$ and
$SU(3)$ groups are the most common, representing spin, flavor, or color degrees of freedom.
Matrix models, on the other hand, involve systems invariant under a larger $SU(N)$ symmetry, and
eventually $N$ is taken to infinity to achieve a particular scaling limit.

Independently, magnetic systems with $SU(N)$ symmetry have been considered in the context of ultracold
atoms \cite{Ghu,Gor,Zha,Mag,Cap}, spin chains \cite{Aff,Pola},
or of interacting atoms on lattice sites \cite{KT,BSL,RoLa,YSMOF,TK,Totsuka,TK2} and in the presence of $SU(N)$ magnetic fields \cite{DY,YM,HM}.
Such systems consist of a large number of components ("atoms"), each carrying an irreducible representation
of the symmetry group, and the full system transforms under the same symmetry in the direct sum of the representations of the components. The decomposition of the states of the full system into
irreducible representations (irreps) of the symmetry group is, then, of physical relevance.

In previous work we derived the statistics of such decompositions \cite{Polychronakos:2023yhq} and investigated the properties
of $n$ coupled $SU(N)$ atoms in the ferromagnetic regime, that is, in the regime where the mutual interaction
of $SU(N)$ components would tend to "align" their charges, also in the presence of an external nonabelian
magnetic field coupled to the system \cite{Phases}. We studied the system in its thermodynamic limit of a large
number of atoms $n \gg 1$ and uncovered a rich phase structure, qualitatively and quantitatively different from that of usual
$SU(2)$ ferromagnets, involving various critical temperatures, hysteresis effects, coexistence of phases,
and latent heat transfer during phase transitions.

The above raise the obvious question of the properties of such systems when both the large-$n$
(thermodynamic) and large-$N$ limits are taken. This is the topic of the present work.
We consider the number of irreducible components in the decomposition of $n$ fundamental irreps of $SU(N)$
weighted by a (positive) power of their dimension, which can be viewed as the infinite-temperature partition
function of an exotic ferromagnet. We derive the large-$(N,n)$ properties of this quantity in the appropriate
scaling in that limit, which requires $n/N^2 \sim 1$ as $N,n \gg 1$, and demonstrate that, in that limit, the
representation content of the system undergoes a phase transition in this parameter.
This transition is model-independent, as the Hamiltonian becomes irrelevant in the infinite-temperature limit,
and thus represents a universal feature of such systems at high temperatures.
Although the large-$N$ limit may appear somewhat unnatural for physical applications, the weaker scaling
of $N \sim \sqrt n$ required for this limit allows it to be achieved for reasonable values of $N$ even for large
systems.

In the following sections, we review the relevant mathematics of decomposing a large number of
fundamental $SU(N)$ representations into irreducible components, stressing the fermion momentum
parametrization of irreps and related duality properties. Subsequently, we consider the multiplicity
of each irrep
weighted by various powers of the number of its states (dimension) in the large-($N,n$) limit and solve for
the saddle point irrep that maximizes it. We uncover a {\it fourth-order} phase transition in the order
parameter $t=n/(4N^2)$ in the generic situation, which becomes a stronger, third-order transition between
a duality preserving and a duality breaking phase for the unweighted multiplicity,
and disappears altogether when weighting by the number of states. We will conclude with some remarks
on possible applications and directions for further investigation.

\section{Fermions and $SU(N)$ group theory: A brief review}
\label{generalities}

We review here the description of irreducible representations of $SU(N)$ as
$N$-fermion energy eigenstates on the circle \cite{MiPo} and the corresponding composition rules in the
fermion picture focusing on the results relevant for our considerations.

\no
The correspondence of irreps of $SU(N)$ and $N$-fermion energy eigenstates is most readily
established by considering the action of a particle moving freely on the group manifold $U(N)$ \cite{Polb} with Lagrangian
\be
{\cal L} = -\half \tr \left( U^{-1} {\dot U} \right)^2\ ,
\label{Lsun}
\ee
where $U$ is an $N$-dimensional unitary matrix and overdot signifies time derivative. Classically, the particle
performs geodesic motion on $U(N)$. Quantum mechanically, energy eigenstates are matrix elements of
irreps of $U(N)$, the energy corresponding to the quadratic Casimir of
the irrep. 

The Lagrangian \eqn{Lsun} is invariant under unitary conjugations of $U$ (as well as left- and
right-multiplications by constant unitary matrices), with a conserved generator $P = i [U^{-1} , {\dot U}]$,
so we can impose the constraint that this conserved generator vanishes. Quantum mechanically,
this amounts to
choosing states invariant under unitary conjugations of $U$. Energy eigenstates in this subspace
become the conjugation-invariant linear combinations $\sum_a r_{aa} (U) = \tr r (U) = \chi_r (U)$,
the latter being the character of the irrep, and depend only on the eigenvalues of the matrix $U$.

Classically, we can implement the conjugation invariance constraint at the Lagrangian level by setting
\be
U(t) = V\, \diag \{ z_j \}\, V^{-1} \, ,\quad z_j := e^{i x_j} \ ,\qq  x_j \equiv x_j + 2\pi .
\ee
The vanishing of $P$ implies that $V$ can be chosen time idependent,
and (\ref{Lsun}) becomes the Lagrangian of $N$ free particles on the unit circle with coordinates
$x_j$. Since exchanging the eigenvalues is a special case of unitary conjugation, states
$\phi (x_1,\dots,x_N)$ are invariant under exchange of the $x_j$ and are, in principle, bosonic.
However, upon quantization, the change of variables from $U$ to $x_j$ introduces in the measure the
absolute square of the Vandermonde determinant $|\Delta (\bz)|^2$, with
\be
\label{vanderm}
\Delta (\bz) = (z_1 \cdots z_N )^{-{N-1 \over 2}} \left|
\begin{array}{ccccccccc}
z_1^{N-1} & z_1^{N-2} & \cdots & z_1  & 1 \\
z_2^{N-1} & z_2^{N-2} & \cdots & z_2  & 1 \\
\vdots & \vdots & \ddots & \vdots & \vdots  \\
z_N^{N-1} & z_N^{N-2} & \cdots & z_N  & 1 \\
\end{array}
\right|
\ .
\ee
(boldface $\bz$ stands for the full set of variable $z_1, \dots , z_N$, and similarly for
other sets of $N$ quantities).
The prefactor $ (z_1 \cdots z_N )^{-{N-1 \over 2}}$ is a pure phase introduced to make
$\Delta (\bz)$ invariant under uniform shifts of the $x_i$'s.
Upon incorporating one factor $\Delta (\bz)$ into the wavefunction
\be
\psi (\bz) = \Delta (\bz)\, \phi (\bz)\ ,
\label{psen}\ee
the integration measure in the $x_j$ becomes flat and
the Hamiltonian becomes the standard free $N$-particle Hamiltonian on the $x_j$. Because
of the prefactor $\Delta (\bz)$ the states $\psi (\bz)$ are antisymmetric upon 
interchanging any two of the $x_i$. This establishes
the correspondence between the conjugation-invariant sector of the model (\ref{Lsun}) and free fermions on
the circle, and thus of irreps of $SU(N)$ and free fermion energy eigenstates. In particular, the fermion
energy eigenstate corresponding to an irrep $r$ with energy $E = C_2 (r)$ in terms of the corresponding
quadratic Casimir becomes
\be
\psi_r (\bz) = \Delta (\bz) \,\chi_r (\bz)\ ,
\label{psar}\ee
with $\chi_r (\bz)$ the character of irrep $r$ in terms of the eigenvalues $z_j$ of $U$.

\no
The single-particle
spectrum on the circle consists of discrete momentum eigenstates with eigenvalue 
$k = 0,\pm 1,\pm 2,\dots$ and energy $E_k = k^2 /2$. An $N$-fermion energy eigenstate
corresponds to filling $n$ of the single-particle states with fermions. Call $k_1 > k_2 > \dots > k_N$
the momenta of these states in decreasing order.
The (unnormalized) wavefunction corresponding to this state is given by the Slater determinant
\be
\psi_\bk (\bz) = \left|
\begin{array}{ccccccccc}
z_1^{k_1} & z_1^{k_2} & \cdots & z_1^{k_{N-1}}  & z_1^{k_N} \\
z_2^{k_1} & z_2^{k_2} & \cdots & z_2^{k_{N-1}}  & z_2^{k_N} \\
\vdots & \vdots & \ddots & \vdots & \vdots  \\
z_N^{k_1} & z_N^{k_2} & \cdots & z_N^{k_{N-1}} & z_N^{k_N} 
\end{array}
\right|\ .
\label{slater}
\ee
The total momentum of the fermions $K = k_1 + \cdots + k_n$ corresponds to the
states picking up a phase $e^{i c K}$ upon the shift $x_j \to x_j +c$, that is, upon $U \to e^{ic} U$.
It thus represents the $U(1)$ charge of the state. We may shift all momenta by a constant,
changing $k$ and the $U(1)$ charge without affecting the $SU(N)$ part of the
states, which  can then be labeled by the $N\hm1$ shift-invariant integers
$k_1 -k_N > k_2 -k_N > \dots > k_{N-1} - k_N >0$.
Alternatively, we can neutralize the $U(1)$ charge by introducing the prefactor $(z_1 \dots
z_N)^{-\sum_i k_i /N}$ in \eqn{slater}, similarly to the prefactor introduced in $\Delta ({\bf z})$.
With this additional prefactor, \eqn{slater}  maps to \eqn{vanderm} for the
singlet representation for which $k_i = N-i\,,  i=1, \dots ,N$.

\noindent
Finally, we note the correspondence with the standard Young tableau\footnote{In standard physics convention,
we use the term tableau instead of the term diagram that is used in mathematics, where tableau stands for
a diagram with properly ordered integer entries in its boxes.}. This is done by expressing $k_j -k_N$ 
in terms of variables $\ell_j$ as
\be
\label{ellk}
\ell_j = k_j - k_N + j-N\ , \qq \ell_1 \geqslant \ell_2\geqslant\dots \geqslant \ell_{N-1} \geqslant 0\ .
\ee
The non-negative, ordered integers $\ell_j$ represent the length of rows $j=1,2,\dots , N\hm1$ of the
Young tableau of the irrep corresponding to the fermionic state. 
The transition from $k_j$ to $\ell_j$ is, in fact, bosonization (in the sense that any two consecutive $\ell$, unlike the $k$, can be equal), the $\ell_j$ corresponding to the possible
momenta of $N$ bosons on the circle.

\subsection{Composition of (fundamental) representations}

Consider the direct product $r_1 \times r_2$ of two (possibly reducible) representations $r_1$ and $r_2$. The basic relation
\be
 \chi_{r_1\times r_2} (U)= \tr\, (r_1 \times r_2 ) (U) = \tr\, r_1 (U)\, \tr\, r_2 (U) = \chi_{r_1} (U)\, \chi_{r_2} (U)\ ,
\ee
implies, through  (\ref{psen} and \ref{psar}), that their corresponding fermion states are related as
\be
\label{psis}
\psi_{r_1 \times r_2} (\bz) =\Delta(\bz) \chi_{r_1} (\bz)\chi_{r_2} (\bz)=
 {\psi_{r_1} (\bz)\, \psi_{r_2} (\bz) \over \Delta(\bz)}
= \psi_{r_1} (\bz)\, \chi_{r_2} (\bz) = \psi_{r_2} (\bz)\, \chi_{r_1} (\bz)\ .
\ee
 Here we need the case of the fundamental $f$
for which the  character is simply
\be
\label{lowreps}
\chi_f (\bz) = \sum_{i=1}^N z_i \ .
\ee
The composition of two irreps $r \otimes f$, the second one being the fundamental, is then straightforward. 
The fermion wavefunction of the state corresponding to $r \otimes f$ is, using \eqn{psis} and \eqn{lowreps}, simply
\be
\psi_{r \otimes f} (\bz) = \psi_{r} (\bz)\chi_f (\bz)= \psi_{r} (\bz) \sum_{i=1}^N z_i\ .
\ee
This can be used to obtain the fermionic state corresponding to the composition of several
($n$ in number) fundamental irreps $f$. The original, singlet state is simply $\Delta (\bz)$ and
an iteration of the above formula yields
\be
\psi_{N,n}(\bz)= 
\Delta (\bz) \Big(\sum_{i=1}^N z_i \Big)^n :=
\sum_{\bk}\, d_{n;{\bf k}}\, \prod_{i=1}^N z_i^{k_i} \ .
\label{psiNn}
\ee
Since $\psi_{N,n}$ is antisymmetric in the $z_i$, the coefficients  $d_{n;{\bf k}}$ appearing
above are fully antisymmetric in the $k_i$. When the $k_i$ are in decreasing order,
$d_{n;{\bf k}}$ gives the multiplicity of the irrep labeled by $k_1 > \dots > k_N$.

\no
To derive an explicit combinatorial expression for the multiplicity we first focus on
the coefficients produced by the term $(z_1+\cdots+z_N)^n$, denoted by $D_{n;{\bf k}}$.
We have
\be
\left(\sum_{i=1}^N z_i \right)^n := \sum_{k_1,\dots,k_N} D_{n;{\bf k}} \prod_{i=1}^N z_i^{k_i} \ ,\qq
D_{n;{\bf k}}=
 \delta_{k_1 + \cdots +k_N ,n} ~ {n! \over \prod_{i=1}^N k_i !} \ , \qq  k_i \geqslant 0\ .
\label{rann}
\ee
Incorporating the Vandermonde factor in (\ref{psiNn}) we eventually obtain \cite{Polychronakos:2023yhq}
\be
\label{lhq1}
d_{n;{\bf k}}= 
n!\, {\displaystyle \prod_{j>i=1}^N (k_i - k_j) \over \displaystyle \prod_{i=1}^N k_i !}=
n!\, {\displaystyle \Delta (\bk) \over \prod_{i=1}^N k_i !}\ 
\quad\text{with}\quad \sum_{i=1}^N k_i = n+{N(N-1)\over 2}\ .
\ee
The dimension of the irrep expressed in terms of the $k_i$'s becomes
\be
\label{dsk}
\dim(\bk) = \prod_{j>i=1}^N {k_i - k_j  \over j-i} = {\Delta(\bk) \over \prod_{s=1}^{N-1} s!}\ .
\ee

\subsection{Momentum density and a group duality}

We conclude by giving a "second quantized" expression for the $d_{n,k_1,\dots,k_N}$ that is useful
in the large-$N,n$ limit. Thinking of the $k_i$ as a distribution of fermions on the positive momentum
lattice $s=0,1,\dots$, we define the discrete momentum density of fermions $\rho_s$ equal to one on points
$s$ of the momentum lattice where there is a fermion and zero elsewhere, that is,
\be
\label{rs}
\rho_s = \sum_{i=1}^N \delta_{s,k_i}\ .
\ee
Clearly $\rho_s$, and in accordance with \eqn{lhq1}, satisfies the relations
\be
\sum_{s=0}^M \rho_s = N\ ,\qq \sum_{s=0}^M s\, \rho_s = K = n+{N(N-1) \over 2}\ ,
\label{rNn}
\ee
where $M$ is a cutoff momentum that can be chosen arbitrarily as long as it is bigger than all the $k_i$'s.
Then, it can be easily seen that  (\ref{lhq1}) can be written as \cite{Polychronakos:2023yhq}
\be
d_{n,\bk} = n!\, \prod_{t>s=0}^M (t-s)^{(\rho_s -1) \rho_t}\ .
\label{discro}
\ee

\no
The integer $M$ could in principle be taken to infinity. However, keeping it finite serves to demonstrate an interesting
particle-hole duality of the formulae. Define
\be
{\tilde \rho}_s = 1-\rho_{M-s} \ , \quad s=0,1,\dots,M\ .
\ee
Clearly ${\tilde \rho}_s$ is the density of holes on the lattice $[0,M]$ with the momentum reversed.
Moreover, using \eqn{rNn}, ${\tilde \rho}_s$ satisfies
\be
\sum_{s=0}^M {\tilde \rho}_s = M-N+1\ , \qquad  \sum_{s=0}^M s\, {\tilde \rho}_s 
= n+{(M-N+1)(M-N) \over 2}\ .
\ee
Therefore, ${\tilde \rho}_s$ represents an irrep of $SU(M-N+1)$ with the same excitation $n$ (total
number of boxes) but with the
rows in the Young tableau of the $SU(N)$ irrep turned into columns for $SU(M-N+1)$, which defines the
dual irrep. One can check that
\be
d_{n,\bk} = n!\, \prod_{t>s=0}^M (t-s)^{({\tilde \rho}_s -1) {\tilde \rho}_t}\ .
\label{dualrho}
\ee
That is, in the decomposition of the tensor product of $n$ fundamentals of $SU(N)$, the multiplicity
of any given irrep is the same as the one for its dual irrep in the product of $n$ fundamentals of $SU(M-N+1)$.
Note that this relation holds for any $M$ such that $M \geqslant k_1 > k_2 > \cdots > k_N$.

\no
This duality between $SU(N)$ and $SU(M-N+1)$ can be turned into a self-duality if we choose $M=2N-1$,
which is possible if $k_1 \leqslant 2N-1$, that is, $\ell_1\leqslant N$. This will be guaranteed in the case
$n \leqslant N$ since then, indeed, $\ell_1\leqslant n\leqslant N$.

\section{Large-$N,n$ limits}

The two parameters at our disposal in the $SU(N)$ case, namely $N$ and $n$, can be taken to be large in
various ways, leading to different large-$N,n$ limit regimes. The dimensionality of the Hilbert space of the
system is $N^n$, and both the $n\gg 1$ and the $N \gg 1$ limits are driving it to infinity, although in qualitatively
different ways: the limit $n \gg 1$ can be viewed as a standard thermodynamic limit increasing the number of
individual components (fundamental irreps) of the system, while the $N \gg 1$ limit swells the available phase
space per degree of freedom. The relative scaling between $n$ and $N$ becomes important and, as we shall see,
the distribution of irreps undergoes a phase transition at some critical line $n \sim N^2$ in the two-dimensional parameter space spanned by $N$ and $n$. The exact critical line is fixed by the statistical quantity of interest in the
problem, namely, number of irreps, number of states, or a more general combination.

\no
In the following we will analyze the large-$n,N$ limit of the distribution of irreps, derive the
dominant distribution, establish the existence of a phase transition, and determine the order of the transition.

\subsection{$n\gg N \sim 1$}

The thermodynamic limit $n \gg N \sim 1$ is the most straightforward. It was derived in 
\cite{Polychronakos:2023yhq} and used
in \cite{Phases} to analyze the $SU(N)$ ferromagnet and determine its intricate phase transition diagram.
We can use the Stirling approximation in the combinatorial formula for $d_{n;\bk}$
given in \eqn{lhq1}. In that limit, $k_i\gg 1$ since from \eqn{lhq1} their sum is of order $n$.
The result is 
\be
d(N,n;\bk) = \prod_{s=1}^{N-1} s! \,{N^{n+N^2/2}\, e^{N^2(N^2 -1) \over 24n} \over \sqrt{2\pi}^{N-1} n^{(N^2 -1)/2}}
\, \dim(\bk) \, e^{-{N\over n} c_2 (\bk) }\ ,
\label{dirr}
\ee
with the dimension of the irrep and the corresponding quadratic Casimir given by \eqn{lhq1} and
\be
c_2 (\bk) = {1\over 2} \sum_{i=1}^N \left( k_i - {K\over N}\right)^{\bb 2} - {N(N^2 -1) \over 24}\ ,
\ee
where factors subleading in $1/n$ are eliminated. Distribution \eqn{dirr} implies that in the limit $n\gg 1$
the deviations of $k_i$ from their mean value $K/N \sim n/N$ scale as
\be
k_i - {  K \over N} \sim \sqrt{n}\ .
\ee

\subsection{$N,n \gg 1$}

The limit where both $N$ and $n$ are large is more interesting. It that limit it does not make sense any more
to define a continuous distribution
$d(N,k_1,\dots,k_N)$, as the dimensionality of the space of $k_i$ grows to infinity for large $N$.
Instead, it is possible to define a {\it density of points} $\rho(k)$ that is the continuous
version of $\rho_s$ defined in \eqn{rs} smoothed over the position of lattice points $s$ around
momentum $k$,
and express the number of irreps $d_{n,\bk}$ as a functional of this density $\rho(k)$. Alternatively, $\rho (k)$ can be defined through the continuous momentum function
\be
k(j) \simeq k_j ~~ (j=1,\dots,N)\ ,\qq {\rho(k) = -{dj \over dk(j)} }~~~\text{at}~~ k(j) = k 
\ee
(the minus sign in the definition  of $\r(k)$ is needed to ensure a positive $\r$ since $k_j$ is decreasing with $j$).
Then, using the continuous version of (\ref{discro}), the logarithm of the number of irreps
$d_{n,\bf k} := d_n [\rho(k)]$ becomes in the continuous approximation
\be
\ln {d_n [\rho(k)]\ov n!} =
 \int_0^\infty \bb dk \int_0^k \bb dk' \, \rho(k) \rho(k') \ln(k - k' ) 
- \int_0^\infty \bb dk \int_0^k dk' \rho(k)\, \ln(k-k')\, ,
\label{lndn}
\ee
up to a constant of ${\cal O}(1)$ and where we have moved the $\r$-independent term $\ln n! $ to 
the left hand side. Rewriting the double integral in a way symmetric in
$k$ and $k'$ and performing the integral over $k'$ in the second term, we obtain 
\be
\ln {d_n [\rho(k)]\ov n!} =
 \ha \int_0^\infty \bb dk \int_0^\infty  \bb dk' \, \rho(k) \rho(k') \ln |k - k' | 
- \int_0^\infty \bb dk\, \rho(k)\, k (\ln k -1)\ .
\label{lndn2}
\ee
Integrals
involving singular kernels, such as $\ln|k-k'|$, are always defined via their principal value, since the discrete
version omits the points $k_i = k_j$ while including points $k_i = k_j \pm 1$, leading to a symmetric
regularization. Similarly, the logarithm of the dimension of the irrep \eqn{dsk} becomes in the continuum limit
\be
\ln \dim [\rho(k)] =-\ln {\displaystyle{\prod_{s=1}^{N-1} s!}}
 + \int_0^\infty \bb dk \int_0^k \bb dk' \, \rho(k) \rho(k') \ln(k - k' ) \ .
\label{dimr}
\ee
The distribution $\rho(k)$ satisfies the constraints
\be
0\leqslant  \rho(k) \leqslant 1 \ ,\qq \int_0^\infty dk\, \rho (k) = N\ ,\qq  \int_0^\infty dk \, k\, \rho(k) = {N^2 \over 2} +n\ .
\label{constr}
\ee
The first constraint arises from the fermionic property $k_i \geqslant k_{i+1}+1$, while the other two are
the continuous, large-$N$ version of (\ref{rNn}). 

In the large-$n,N$ limit the statistics of irreps will be will be dominated by one particular distribution
$\rho(k)$ corresponding to one irrep, the contribution from other irreps falling off exponentially
as the density deviates from that distribution. The determination of the dominant irrep depends on the
quantity of interest. In a pure mathematical context, the number of irreps $d_{n;{\bf k}}$ could be
the quantity of interest, and we would need to maximize it with respect to $\rho(k)$. That is, we would
maximize the expression in \eqn{lndn} under the constraints \eqn{constr}.
In statistical physics applications, on the other hand, the relevant quantity is the total number of states at
given energy and other thermodynamic state variables. Assuming that the irrep determines all such
variables (energy etc.), the relevant quantity is the total number of a given irrep times its dimensionality
(number of states), that is,
\be
\label{mw2}
m_{n;{\bf k}} = \dim(\bk) d_{n;\bf k} = 
{n!\, \Delta({\bf k})^2 \over \,\displaystyle{\prod_{s=1}^{N-1} s!  \prod_{i=1}^N k_i !}\,}\ ,
\quad\text{with}\quad {\sum_{i=1}^N k_i }= n+{N(N-1)\over 2}\ ,
\ee
where we used (\ref{lhq1} and \ref{dsk}). In terms of the (discrete) fermion density $\rho_s$ we have that
\be
m_{n;\rho} = {n!\over \, \prod_{s=1}^{N-1} s!} \,
\prod_{t>s=0}^M (t-s)^{({2\rho}_s -1) {\rho}_t}\ .
\label{dmdiscr}\ee
The above obeys the formal duality invariance $\rho_s \to {\tilde \rho}_s = \half - \rho_{M-s}$.
However, this is not a true duality since, given that $ \rho_s = 0$ or $1 $, ${\tilde \rho}_s$ takes the 
values $\pm \half$. In the large-$n,N$ limit in which $\rho_s \to \rho (k)$ we obtain from (\ref{lndn} and \ref{dimr}) that
\be
\ln m_n [\rho(k)] =
2 \int_0^\infty \bb dk \int_0^k \bb dk' \, \rho(k) \rho(k') \ln(k - k' ) 
- \int_0^\infty \bb dk\, \rho(k)\, k (\ln k -1)\ ,
\ee
up to a $\rho$-independent constant. This is identical in form to
$\ln (d_n [\rho(k)]/ n!) $ in \eqn{lndn}, the only difference being the factor of $2$
in front of the double integral. 

\no
We can thus consider the general form
\be
S_{w,n} [\rho(k)] = 
 {w\ov 2} \int_0^M \bb dk \int_0^M  \bb dk' \, \rho(k) \rho(k') \ln |k - k' | 
- \int_0^M \bb dk\, \rho(k)\, k (\ln k -1)\ ,
\label{Sw}
\ee
where we ignored any overall $\rho$-independent constant and introduced an upper cutoff $M$
for the $k$ integrals, with the understanding that $\rho(k) = 0$ for $k>M$.
This reproduces the cases $\ln d_n [\rho]$
for $w = 1$ and $\ln m_n [\rho(k)]$ for $w=2$, but can describe a more general situation.
Note that the case of general $w$ would correspond to starting, instead of \eqn{mw2}, with
\be
\label{mww}
m_{w,n;{\bf k}} =\big(\dim(\bk)\big)^{w-1} d_{n;\bf k} = 
{n!\, \big(\Delta({\bf k})\big)^w \over \,\displaystyle{\Big(\prod_{s=1}^{N-1} s! \Big)^{w-1} \prod_{i=1}^N k_i !}\,}\ ,
\qquad {\sum_{i=1}^N k_i }= n+{N(N-1)\over 2}\ .
\ee
There is a clear distinction between the cases $w>1$ and $w<1$. The latter one is rather exotic, and
perhaps unphysical, as  it would correspond to a statistical model with entropy decreasing as the dimensionality
of the irrep increases.

\no
The quantity $S_{w,n} [\rho(k)]$ is invariant under the formal duality transformation
\be
\rho(k) \to {\tilde \rho}(k) = w^{-1} - \rho(M-k)\ .
\label{dualw0}
\ee
analogous to the one for the $w=2$ case \eqn{dmdiscr}. For
$\tilde \rho$ to obey the fermionic constraint $0<{\tilde \rho}<1$, $\rho$ must satisfy
\be
w^{-1} -1 < \rho < w^{-1}\ ,
\label{dres}\ee
which implies that ${\tilde \rho}$ will also satisfy it. Further, ${\tilde \rho}$ satisfies the integral
constraints \eqn{constr} with modified parameters ${\tilde N}$ and ${\tilde n}$
\be
{\tilde N} = {M\over w} - N \ ,\qq  {\tilde n} = n+{(w-1)M(M-2wN) \over 2 w^2} \ .
\ee
For the special value $M= 2 w N$ we see that ${\tilde N} = N$ and ${\tilde n} = n$ and therefore
the transformation \eqn{dualw0} becomes a self-duality.
Note, however, that this duality holds only for densities satisfying \eqn{dres} so it
remains a restricted invariance. Its domain, in particular, does not include the singlet.

To calculate the dominant irrep, that is, the distribution $\rho(k)$ maximizing $S_{w,n}$, we maximize
$S_{w,n} [\rho(k)]$ while enforcing the constrains (\ref{constr}) via two Lagrange multipliers.
That is, we extremize
\be
 S_{w,n} [\rho(k)] -\mu \left(\int_0^\infty dk\, \rho(k) - N \right) -\lambda
\left( \int_0^\infty dk\, k\, \rho(k) - n - {N^2\over 2}\right)\ .
\label{r1}
\ee
where from now on we take $M\to \infty$ keeping in mind that, in general, $\r(k)$ will vanish outside a
finite range. Using (\ref{lndn}) and setting the functional derivative with respect to $\rho(k)$ to zero yields
\be
w \int_0^\infty dk'\, \rho(k') \ln|k-k'| = k(\ln k -1) + \mu +\lambda k\ .
\label{r2}
\ee
Further differentiating \eqn{r2} with respect to $k$ we obtain
\be
w\int_0^\infty dk' {\rho(k')\over k - k'} = \ln k + \lambda\ .
\label{eomr}
\ee
The above equation must hold for $k$ such that $\rho (k) \neq 0$ and $\rho (k) \neq 1$, since in
empty regions with $\rho (k) = 0$ there are no $k_i$ to vary, and in fully filled ones with $\rho (k) =1$
the $k_i$ cannot vary.

\begin{figure} [th!]
\begin{center}
\includegraphics[height= 6.5 cm, angle=0]{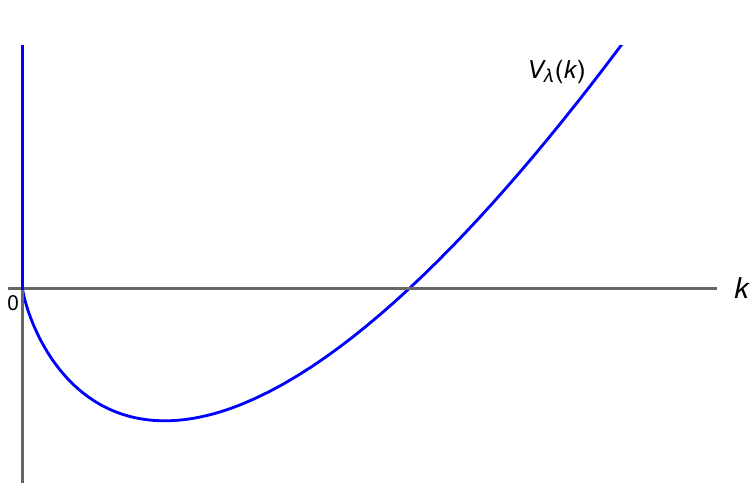}
\end{center}
\vskip -.5 cm
\caption{\small{The potential $V_\lambda (k)$ for a generic value of $\lambda$. It has a "rigid wall" at $k=0$
and forms a well for all values of $\lambda$, allowing for a distribution $\r(k)$ localized inside the well,
either touching $k=0$ or with support entirely at $k>0$.}}
\label{PotentialVk}
\end{figure}
\no
Hence, from (\ref{r2}) we see that the problem amounts to finding the equilibrium distribution of a
large number of particles repelling each other with a logarithmic potential of strength $w$, 
inside an external potential given by the right hand side of (\ref{r2}), that is 
\be
\label{potvl}
V_\l(k)=k(\ln k-1) + \lambda k\ ,
\ee
where we have omitted the constant term $\mu$. This potential is depicted in fig. \ref{PotentialVk}.
It has a rigid "wall" at
$k=0$ and goes to infinity as $k\to \infty$. Therefore, it may always support a finite lump of particles, spread around its minimum value $V_{\l,{\rm min}} = -e^{-\l}$ arising at  $k= e^{-\lambda}$.
The issue is whether this lump extends all the way to the boundary at $k=0$, and if it respects the
condition $0\leqslant \rho(k) \leqslant 1$. As we shall see, this depends on the values of $N,n$
and $w$.

In the following sections, we solve the minimization problem and obtain the dominant $\rho (k)$. We
will first treat the case $w=1$ (maximizing $d_n [\rho (k)]$), since the solution simplifies and has
some special properties, and then extend it to general $w$.

\subsection{$w=1$, maximal $d_n$}

In this case the equation satisfied by $\rho (k)$ is simply \eqn{eomr} with $w=1$
\be
\int_0^\infty dk' {\rho(k')\over k - k'} = \ln k + \lambda\ ,
\label{eomr1}\ee
together with the constraints \eqn{constr}. Moreover, for $w=1$ the duality relation \eqn{dualw0}
becomes exact,
as it preserves the range of $\rho$ and maps the singlet to itself. Depending on whether the inequality
constraint $\rho \leqslant 1$ is saturated in a finite domain, we distinguish two cases, corresponding to
broken or unbroken duality symmetry.

\subsubsection{Duality breaking phase $n>N^2 /4$}

We start by assuming that the distribution $\rho(k)$ does not reach the "wall" on the left at $k=0$, i.e.,
it is nonzero inside an interval $0<a<k<b$ and vanishes outside. Then solving (\ref{eomr})
becomes a standard single-cut Cauchy problem. We define the resolvent
\be
\label{ukr1}
u(z) = \int    dk\, {\rho(k) \over z-k}\ ,
\ee
with $z$ on the upper complex plane. Its real and imaginary part on the real axis reproduce $\rho(k)$ and its
Hilbert transform
\be
u(k+i\epsilon) = \Xint{-}  dk' {\rho(k') \over k-k'} -i\pi  \rho(k)\ .
\label{ukr}
\ee
Therefore, a function that is analytic on the upper half plane and its real part on the real axis equals
$\ln k +\lambda$ will equal $u(z)$ up to an additive constant, and its imaginary part will fix $\rho(k)$.
In standard fashion, we write
\be
u (z) = {1\over 2\pi i} \sqrt{(z-a)(z-b)} \oint ds\, {\ln s +\lambda \over (s-z) \sqrt{(s-a)(s-b)}}\ ,
\label{rescut}
\ee
where the contour winds in the clockwise direction around the cut of the square root but does not
include the
singularity at $z$ and the cut of the logarithm (see fig. \ref{contour1})
\begin{center}
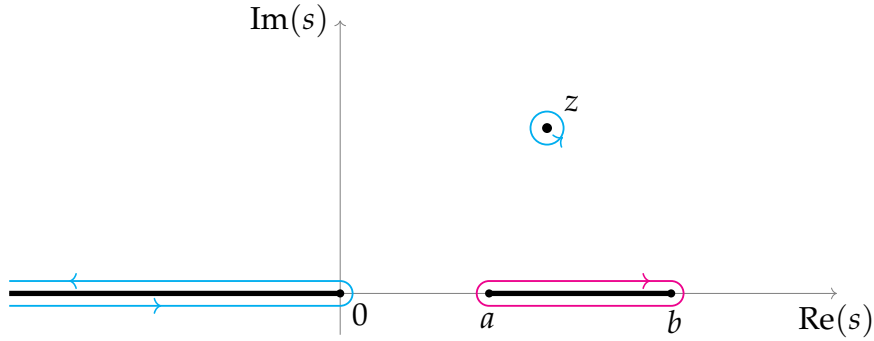

\begin{tikzpicture}[scale=1.1, decoration={markings,
mark=at position 0.9cm with {\arrow[line width=0.6pt]{<}},
mark=at position 7.4cm with {\arrow[line width=0.6pt]{<}}
}
]
\draw[help lines,->] (-4,0) -- (6,0) coordinate (xaxis);
\draw[help lines,->] (0,-0.5) -- (0,3.3) coordinate (yaxis);

\fill (2.5,2.0) circle (1.7pt);\fill (1.8,0) circle (1.4pt);\fill (4,0) circle (1.4pt);\fill (0,0) circle (1.4pt);

\node at (0.24,-0.25) {$0$};
\node at (2.8,2.3) {$z$};
\node at (1.78,-0.33) {$a$};
\node at (4.04,-0.34) {$b$};
\path[draw,line width=0.7pt,postaction=decorate] (2.3,2.0) [cyan] arc(180:-180:0.2);

\path[draw,line width=2.0pt] (1.8,0) 
-- (4,0)
;

\path[draw,line width=0.7pt,postaction=decorate] (4,-0.15) [magenta] arc(-90:90:0.15)
-- (1.8,0.15) arc(90:270:0.15) -- (4,-0.15) ;

\path[draw,line width=2.0pt] (-4,0) -- (0,0);
\path[draw,line width=0.7pt,postaction=decorate] (-4,0.15) [cyan] -- (0,0.15) arc(90:-90:0.15) -- (0,-0.15) -- (-4,-0.15);

\node[below] at (xaxis) {$\text{Re}(s)$};
\node[left] at (yaxis) {$\text{Im}(s)$};

\end{tikzpicture}
\vskip -.4 cm
\captionof{figure}{\small{Contour of integration in the $s$-plane. The original (magenta) contour
around the square root cut on $(a,b)$ is pulled back to the two (cyan) contours around the pole at $z$
and the logarithm cut on $(-\infty,0)$.}
\label{contour1}}
\end{center}

Pulling back the contour we pick up the pole at $s=z$
and the integral around the cut of the logarithm
\be
\begin{split}
u (z) &= \ln z +\lambda -\sqrt{(z-a)(z-b)} \int_0^\infty {ds \over (s+z)\sqrt{(s+a)(s+b)}} 
\\
&= \ln z +\lambda - i \cos^{-1}{2z-a-b \over b-a} + i \cos^{-1}{(a+b)z - 2ab\over (b-a)z}\ .
\end{split}
\label{rescut1}
\ee
For $z =k$ real and between $a$ and $b$ (the region in which $\rho(k)$ does not vanish) 
the last two terms are purely imaginary (the square root factor multiplying the integral provides a
factor $+i$, since we assume that $z$ approaches $k$ from the upper-half complex plane). 
Then, according to (\ref{ukr}), we determine $\rho (k)$ as
\be
\begin{split}
\rho(k) &= {1\over \pi} \cos^{-1}{2k-a-b \over b-a} -{1\over \pi} \cos^{-1}{(a+b)k - 2ab\over (b-a)k}
\\
&= {2\over \pi} \cos^{-1} {\sqrt{k}+\sqrt{ab/k} \over \sqrt{a}+\sqrt{b}}\ ,\qq a\leqslant k\leqslant b\ .
\label{rhocos}
\end{split}
\ee
The last expression above makes clear that $\rho(k)$ indeed vanishes at $k=a$ and $k=b$. It also
makes clear that $\rho (k)$ is positive and never
becomes larger than $1$ (since the argument of $\cos^{-1}$ never becomes negative and thus the
angle does not exceed $\pi/2$), so $\rho (k)$ satisfies the constraint $0\leqslant \rho(k) \leqslant 1$.

The parameters $a$, $b$ and $\lambda$ can be determined by matching the asymptotics of $u(z)$
\be
\label{assym}
\begin{split}
u(z) &=  z^{-1} \int_0^\infty dk\, \rho(k) + z^{-2} \int_0^\infty dk\, k\, \rho(k) + {\cal O}\big(z^{-3}\big)\, 
 \\
&= z^{-1}\, N + z^{-2}\, \bigg(n+{N^2\ov 2}\bigg) +{\cal O}\big(z^{-3}\big)\ , 
\end{split}
\ee
where in the second line we used \eqn{constr}, with those implied from (\ref{rescut1}). We obtain
\be
\label{Nnla}
\begin{split}
& N =  {(\sqrt{b}-\sqrt{a})^2\ov 2}\ ,
\\
&
n+{N^2\over 2} = {2(b-a)^2+(\sqrt{b}-\sqrt{a})^4\over 16} \ ,
\\
&\lambda = -2\ln{\sqrt{a}+\sqrt{b}\over 2}  \ ,
\end{split}
\ee
from which the parameters $a$, $b$ and $\l$ are determined in terms of $N$ and $n$ as
\be
\label{abla}
\begin{split}
&
\sqrt{a}=\sqrt{{N\over 2}}\, \bigg(\sqrt{1+t\ov 2}-1\bigg)\ ,
\\
& \sqrt{b} = \sqrt{{N\over 2}}\, \bigg(\sqrt{1+t\ov 2}+1\bigg)\ , 
\\
&
\lambda = -\ln{N\over 4} -\ln \big(1+t\big) \ ,
\end{split}
\ee
where we defined
\be
\label{torde}
t=4{n\ov N^2}\ .
\ee
The above expressions make clear that, in order for both parameters $n,N$ to remain relevant, their scaling
must be $n \sim N^2$, identifying $t$ in \eqn{torde} as the relevant order parameter. Under that scaling,
$k$, $a$, $b$, and $e^{-\l}$ all scale like $N \sim \sqrt n$.
The form of $\rho (k)$ in \eqn{rhocos} becomes, upon substituting \eqn{abla},
\be
 \r(k) ={2\ov \pi} \cos^{-1} {\sqrt{k\ov N} + {1\ov 4}\sqrt{N\ov k} \big(t- 1\big)\ov \sqrt{t+1}}  \ ,\quad a\leqslant k\leqslant b\ .
\label{rhok}
\ee
It can be checked that  the above density indeed obeys \eqn{constr}.

\no
Since $\sqrt{a} > 0$, \eqn{abla} shows that this solution will exist for $n > N^2 /4$, that is, $t>1$.
For $n < N^2 /4$ the above solution is not valid, and the point $n = N^2 /4$ marks a transition. 
Clearly $\rho(k)$ in (\ref{rhok})
does not satisfy the self-duality condition $\rho(k) = 1-\rho(2N-k)$, so the phase
$n > N^2 /4$ is a duality breaking one.

\subsubsection{Duality preserving phase  $n < N^2 /4$}

For $n = N^2 /4$ ($t=1$) the parameter $a$ is driven to zero and the solution (\ref{rhocos}) of the
previous section becomes
\be
\rho(k) = {1\over \pi} \cos^{-1} {2k-b \over b} = {2\over \pi} \cos^{-1} \sqrt{k \over b}\ ,\qq 0<k<b\ .
\label{rcrit}
\ee
The parameters $b$ and $\lambda$ in this case can be obtained from the $a \to 0$ limit of the corresponding
expressions \eqn{abla} as
\be
\label{bl}
b = 4 \sqrt{n} \ ,\qq \lambda = -\half \ln n\ .
\ee
We note that, now, $\rho(0) = 1$. This marks a transition to a phase where the density $\rho(k)$
saturates to 1 over a finite interval when $n<N^2 /4$.

In fact,  in this phase the expression for the maximal $d_n [\rho(k)]$ develops a
flat region for a range $0\leqslant k \leqslant a$ with the rest of it becoming $N$-independent, for all $n<N^2 /4$.
This is already clear in the case of finite $N>n$: the expression (\ref{lhq1}) for $d_{n,{\bk}}$
also holds if $N$ is replaced by $N' > N$. Specifically, define
the new "extended" momenta $k_i'$ with $i=1,2,\dots,N'$
\be
k_i'= \left\{ \begin{array}{cc}
k_i +N' -N\ , ~~ & i=1,\dots,N \cr
N' - i\ , ~~ & i= N+1, \dots , N'\ .
\end{array} \right.
\ee
identical to the old ones but shifted to the right by $N' - N$ with th "Fermi sea" filled to their left (the
second line above). Then we have
\be
d_{n,{\bf k}} = d_{n,{\bf k'}} .
\ee
This can be shown, e.g., inductively. For $N'=N+1$,
\be
{\displaystyle \prod_{j>i=1}^{N'} (k_i' - k_j') \over \displaystyle \prod_{i=1}^{N'} k_i'!}
={\displaystyle \prod_{j>i=1}^{N} (k_i - k_j) \prod_{i=1}^N (k_i +1) \over 
\displaystyle \prod_{i=1}^{N} (k_i +1)!}
= {\displaystyle \prod_{j>i=1}^{N} (k_i - k_j) \over \displaystyle \prod_{i=1}^{N} k_i!}
\ee
and by induction we can reach any value $N' >N$.
It is clear that
\be
\sum_{i=1}^{N'} k'_i = n+{N' (N' -1)\over 2}\ ,
\label{knn}
\ee
where we used \eqn{lhq1} for $k$.
Therefore, $\{k_i'\}$ represents the same irrep as $\{ k_i\}$ (same Young tableau) arising from the
direct product of $n$ fundamentals of $SU(N')$.
This translates to a corresponding relation in the momentum density description.
It can be checked that the new "extended" density
\be
\rho' (k) = \left\{ \begin{array}{cc}
1\, , ~~ & 0<k<N'-N\, , \cr
\rho(k+N-N')\, , ~~ &k>N'-N\, ,
\end{array} \right.
\label{rhoa}\ee
produces the same $d_n [\rho(k)]$ in (\ref{lndn}).

This provides an extension of the solution (\ref{rcrit}), which was ostensibly valid only for $n=N^2 /4$, to 
any $n > N^2 /4$. 
Specifically, applying (\ref{rhoa}) with $N\to 2\sqrt{a}$ and $N' \to N$ for the configuration (\ref{rcrit}),
and taking also into account that $b=4\sqrt{n}$, we obtain
\begin{figure} [th!]
\begin{center}
\hskip -0.6cm\includegraphics[height= 3 cm, angle=0]{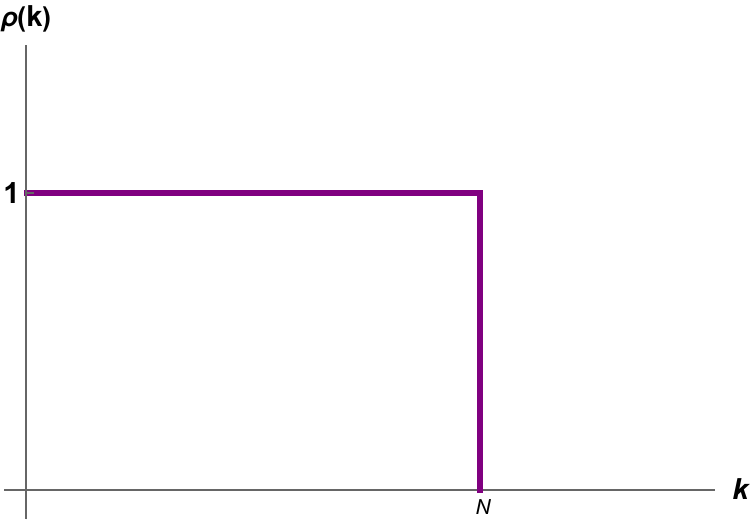}
\hskip .2 cm \includegraphics[height= 3 cm, angle=0]{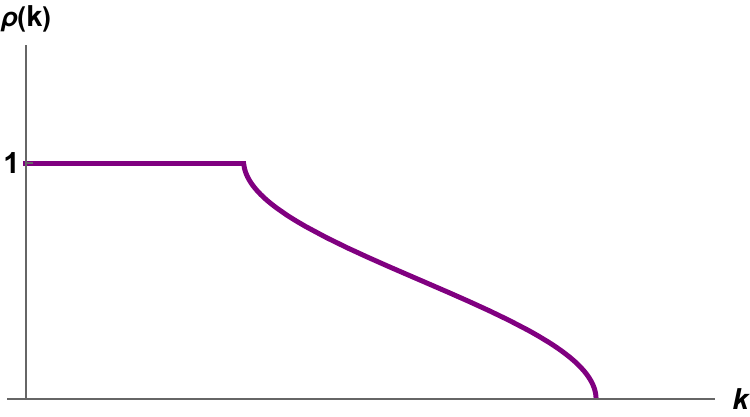}
\hskip .2 cm \includegraphics[height= 3 cm, angle=0]{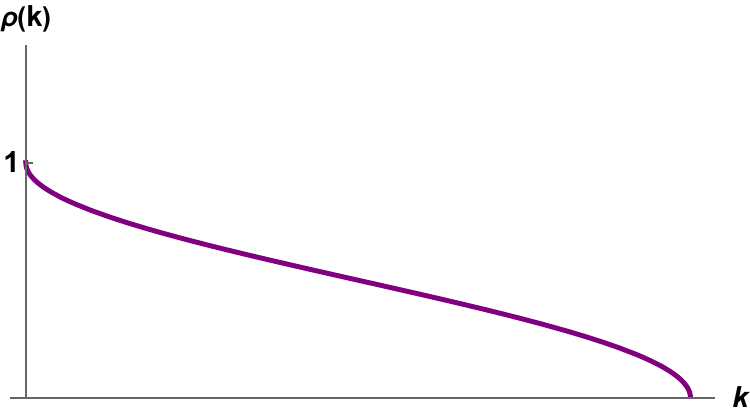} \vskip 0.4cm
\hskip -0.6 cm\includegraphics[height= 3.5 cm, angle=0]{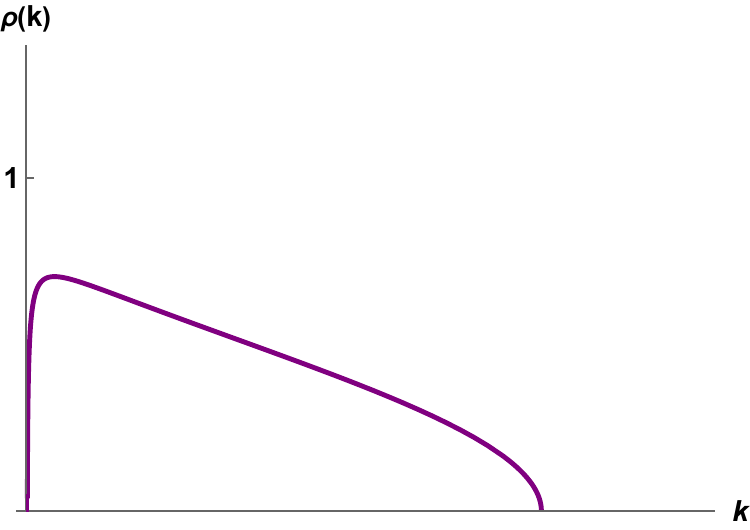}
\hskip .2 cm \includegraphics[height= 3.5 cm, angle=0]{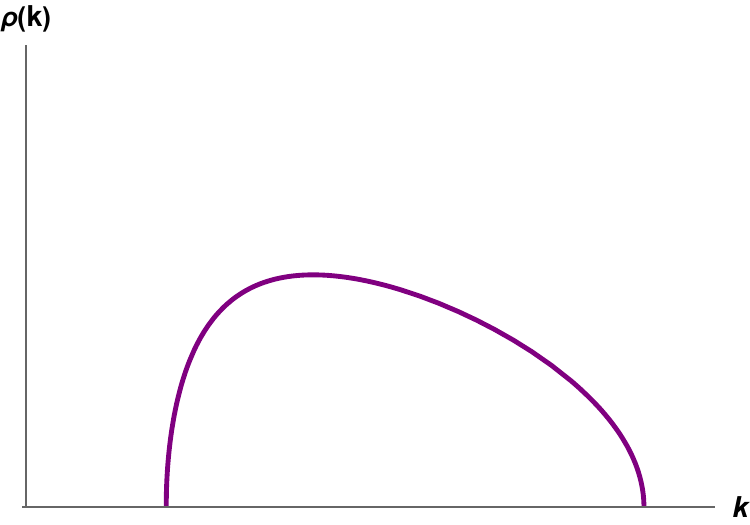}
\hskip 0.2 cm \includegraphics[height= 3.5 cm, angle=0]{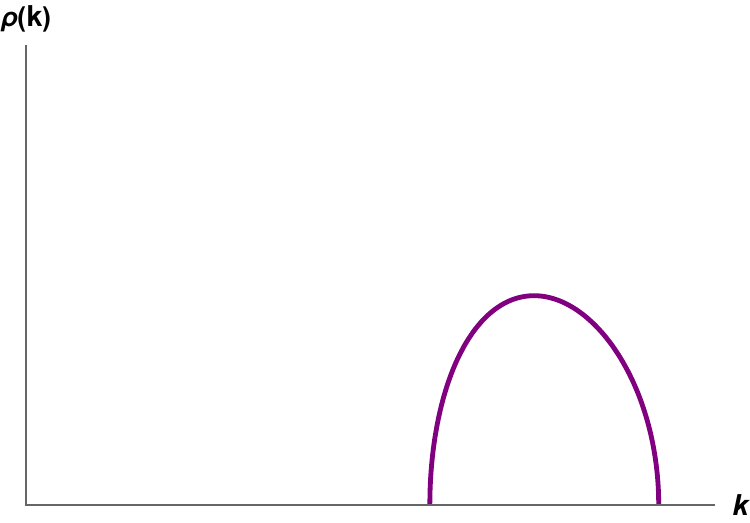}
\end{center}
\vskip -.5 cm
\caption{\small{The distribution $\rho(k)$ for various values of $n/N^2$. For $n=0$ (first panel) the
distribution is a step function corresponding to the singlet. For $0<n<N^2/4$ (second panel) the edge
of the distribution deforms into an inverse cosine. For $n=N^2/4$ (third panel) the deformation reaches
$k=0$, signaling a phase transition. As soon as $n$ exceeds $N^2/4$ (fourth panel) the left edge of the
distribution drops to $\rho(0)=0$, and as $n$ increases (fifth panel) $\r(x)$ has support on a positive interval.
For $n \gg N^2/4$ (sixth panel) it approaches a Wigner semicircle distribution.}}
\label{Rho}
\end{figure}
\be
\label{fjhq2}
\rho(k) =  \left\{ \begin{array}{cc}
1\ ,  ~~ &~~~ 0<k<N-2\sqrt n \ ,
\cr
{\displaystyle{{1\over \pi} \cos^{-1} {k-N \over 2\sqrt{n}}}}\, ,  ~~ &~  N - 2\sqrt{n} <k<  N + 2\sqrt{n}
\end{array} \right.
\ee
and zero elsewhere. 
It can be checked that the above density indeed satisfies \eqn{constr}.
Moreover, it is self-dual, satisfying
\be
\rho(k) = 1-\rho(2N-k) \ ,\qq 0<k<2N\ .
\ee
Therefore, $n < N^2 /4$ represents a duality preserving phase. An evolution of $\rho(k)$ for various values of $n/N^2$
is presented in fig.~\ref{Rho}.

We conclude by pointing out that the distribution \eqn{fjhq2} reproduces the VKLS limiting Young tableau shape
for a large number of boxes $n$ weighted by the Plancherel measure  \cite{VK,LS}.
The VKLS parametrization of Young tableau shapes is achieved by reflecting the Young tableau about its top
row, rotating it by $\pi /4$ in the positive direction to produce a V-based shape, and rescaling it by a factor of
$1/\sqrt{n}$. The coordinates $(x,y)$ of the last box of row $\ell_i$, for large $n$, will be
\be
\ell_i = \sqrt{n\ov  2} (x+y)\ ,\qq  i = \sqrt{n\ov 2} (y-x)
\label{lixy}\ee
Going over to our variables $k_i$, we recall relation \eqn{ellk} between $\ell_i$ and $k_i$. In fact, we will slightly
modify this relation to $k_i= \ell_i + N-i$, so that $\ell_N$ is not necessarily $0$. This has the advantage that the
total number of boxes in the tableau is $n$, at the price of potentially introducing columns of length $N$ in the tableau,
although in the duality preserving phase this will never happen since $k_N =0$.
Then, in the continuum limit $N,n\gg 1$, \eqn{lixy} implies
\be
 k = N +\sqrt {2n}\, x \ .
\ee
The corresponding density of $k_i$ will be
\be
\rho (k) = -{d i \over d k} ={1\over 2}\left(1- {dy \over dx}\right)\ .
\ee
Upon substituting \eqn{fjhq2} and integrating, we obtain
\be
y = \int \bigl(1-2 \rho(N+\sqrt{2 n} x) \bigr) dx = 
\begin{cases}
 {2\over \pi} \left( \sqrt{2-x^2}+ x \sin^{-1} {x\over \sqrt 2}\right)\, ,\quad |x|\leqslant \sqrt{2}\ ,
 \\[6pt]
 |x|\, ,\quad  -{N\ov \sqrt{2n}} \leqslant x \leqslant -\sqrt{2}\quad \text{or}\quad x\geqslant \sqrt{2}\ ,
\end{cases}
\label{VKLS}\ee
where the integration constant was fixed by the condition that $k=0$ for $i=N$. This is the VKLS distribution, with
the left branch of $|x|$ truncated at $x=-N/\sqrt{2n}$, and is valid for all $n < N^2 /4$.
The relation of our work with the Plancherel process and related distributions will be analyzed in an
upcoming publication.

\subsubsection{The Wigner semicircle limit $1\ll N \ll \sqrt n$}

In the limit $n \gg N^2/4 \gg 1$ ($t \gg 1$), deep in the duality breaking region, the particle momenta
$k_i$ lump near the minimum of the effective potential in \eqn{potvl}, where it can be approximated as
a harmonic oscillator. The minimum arises at
\be
k_0 = e^{-\lambda} = {n\over N} + {N\over 4} \simeq {n\over N} = {N  \ov 4}\, t 
\ee
and the potential can be approximated by
\ba
V(k) &\simeq& -{n\over N} + {N\over 2n} \left(k-{n\over N}\right)^2\ .
\label{quadr}
\ea
As a consequence, the distribution $\rho(k)$ will become a Wigner semicircle around $k_0$
with radius $\sqrt{2n}=N\sqrt{t/2}$
\be
\rho(k) = {N\over \pi n} \sqrt{2n - \left(k-{n\over N}\right)^2} ={2\ov \pi} \sqrt{2\ov t} \sqrt{1-2 {(k/N-t/4)^2\ov t}}\ .
\label{Wig}
\ee
The same result is obtained by taking a limit of $\rho(k)$ in (\ref{rhocos}): we set $k/N= {t/ 4} + \sqrt{t}\, y$, where the form of the last term is dictated by the fact that $k$ has range $b-a\simeq N\sqrt{2 t}$ for $t\gg 1$. Expanding $\rho(k)$ for large $t$ we obtain to leading order
\be
\r(k) \simeq {2\ov \pi} \cos^{-1}\biggl ( 1- {1-2 y^2\ov t} \biggr)\simeq {2\ov \pi} \sqrt{2\ov t}\sqrt{1-2 y^2}\ . 
\ee
Reinstating the variable $k/N$ we recover \eqn{Wig}.

\no
Interestingly, this result also derives from the $n\gg N \sim 1$ distribution (\ref{dirr}). The 
exponent of the gaussian exponential corresponds to $k_i$ having a potential $(N/2n) k_i^2$.\,
which is the same potential as (\ref{quadr}).
The Vandermonde factor in front endows the $k_i$ with a logarithmic mutual repulsive potential, while
the delta-function puts the average momentum $(1/N)\sum_i k_i$ to the value $n/N$, leading
to the Wigner semicircle distribution (\ref{Wig}). So the $N \sim 1$ result reliably reproduces the
$1\ll N \ll 2\sqrt{n}$ situation, but not the one for $N \sim 2\sqrt{n}$, which leads to distorted Wigner
semicircle distributions and, eventually, to the phase transition to the self-dual phase.

\subsubsection{Phase transition}\label{transition1}

The point $n=N^2 /4$, or $t=1$, clearly marks a phase transition. To identify the order of the phase
transition and the properties of the two phases we calculate the maximal $d_n$ for the dominant $\rho(k)$
in each phase. We will compute its first few derivatives with respect the order parameter $t$ and will
discover a discontinuity in its third derivative at $t=1$, identifying it as a third-order transition.

We will use the standard result that the derivative with respect to $n$ of any functional $F[\r(k)]$ of
$\r(k)$ that does not explicitly depend on $n$ at its maximum in $\r(k)$ subject to the constraints
\eqn{constr} is given by the Lagrange multipler $\lambda$ for $n$. Indeed,
\ba
\hskip -0.4cm{\del \ov \del n} F[\r(k)]\bb &=&\bb \int_0^\infty dk\, {\d F[\r(k)] \ov \d \r(k) }\,  {\del \r\ov \del n} =\int_0^\infty dk\, (\mu + \lambda k)  {\del \r\ov \del n} \\
\bb&=& \bb \mu {\del \ov \del n}\int_0^\infty dk\, \r + \lambda {\del \ov \del n} \int_0^\infty dk\, k\r 
= \mu  {\del N\ov \del n}  + \lambda  {\del \ov \del n}\Bigl(n + {N^2\ov 2}\Bigr)= \lambda\, ,
\nonumber\ea
where we used the saddle point condition and the constraints. The quantity $\ln (d_n [\rho (k)]/n!)$
that was maximized with respect to $\rho(k)$ indeed does not involve $n$ explicitly, and thus
\be
{\partial \over \partial n}\ln {d_n [\rho (k)]_{\rm max}\over n!} = \lambda\ ,
\label{dern}
\ee
or, in terms of the order parameter $t=4n/N^2$,
\be
{\partial \over \partial t}\ln {d_n [\rho (k)]_{\rm max}\over n!} = {N^2 \over 4} \lambda \ .
\ee
where we have explicitly indicated that it is the density at the maximal configuration.

\no
Before proceeding to compute further derivatives, we may integrate \eqn{dern} above to find 
$d_n [\rho (k)]_{\rm max}$ itself.
In the duality preserving phase $n<N^2 /4$ ($t<1$), for which $\l= -\ha \ln n$ from
\eqn{bl}, we obtain
\be
\label{res1}
\ln {d_n [\rho (k)]_{\rm max}\over n!}  = -{N^2\ov 8} t \Big(\ln t-1+2\ln {N\ov 2}   \Big)=
-{n\ov 2}(\ln n-1) \simeq -\ha \ln {n!}\ ,
\ee
where the integration constant has been fixed such that in the limit $n\to 0$, in which we are left with
the fundamental irrep, $d_n \to 1$.  
Therefore, 
\be
\text{duality preserving phase}:\qq d_{n,{\rm max}} = d_n [\rho (k)]_{\max} \simeq  \sqrt{n!} \ .
\ee
Similarly, for the duality breaking phase, $\l$ is given by \eqn{abla}, so we find 
\be
\label{res2}
\ln {d_n [\rho (k)]_{\rm max}\over n!}  
= 
n-\bigg(n+{N^2\ov 4}\bigg)\ln\bigg({n\ov N}+ {N\ov 4}\bigg)-{N^2\ov 4}\bigg(\ha -\ln{N\ov 2}\bigg)
\ee
where we fixed the integration constant by matching the result at $n=N^2/4$ with the
one in the duality preserving phase, since $\rho(k)$ has no discontinuity at the transition point.
Thus\vskip -0.8cm
\be
\text{duality breaking phase:}~~~~~~d_{n,max} = d_n [\rho_n (k)] = 
N^n {n! {\displaystyle{{\left({N^2\over 4}\right)!}}} \over {\displaystyle{\left(n+{N^2\over 4}\right)!}}}
\left(2 e^{-1/2}\right)^{N^2/4}\ .
\ee
The above was written in a suggestive form, one of several equivalent forms at the $N,n \bb\gg\bb 1$
limit, to display the leading behavior: the multiplicity of the dominant irrep becomes a fraction of the
total number of states $N^n$. As $n$ increases, this fraction becomes
\be
{d_{n,max} \over N^n} \simeq 
\left({N^2\over 4}\right)!\left(2\over n\, e^{1/2}\right)^{N^2/4} \ ,\qq n \gg N^2/4\ .
\ee
It is clear that $\ln d_n/n!$, playing the role of free energy, has no discontinuity at the transition point,
and neither does its first derivative, since $\lambda$ is continuous
across the transition. It turns out that the second derivatives in $t $ for fixed $N$ evaluated at $t=1$ is also continuous. 
Specifically,
\be
\begin{split}
&{\partial \over \partial t}\ln {d_n [\rho (k)]_{\rm max}\over n!}\Big|_{t\to 1^-} =-{N^2 \ov 4}\ln {N\ov 2} = {\partial \over \partial t}\ln {d_n [\rho (k)]_{\rm max}\over n!}\Big|_{t\to 1^+} \ ,
\\
&
{\partial^2 \over \partial t^2}\ln {d_n [\rho (k)]_{\rm max}\over n!}\Big|_{t\to 1^-} =-{N^2 \ov 8}
 = {\partial^2 \over \partial t^2}\ln {d_n [\rho (k)]_{\rm max}\over n!}\Big|_{t\to 1^+} \ ,
\end{split}
\ee
However, the third derivative is discontinuous:
\be
\begin{split}
&{\partial^3 \over \partial t^3}\ln {d_n [\rho (k)]_{\rm max}\over n!}\Big|_{t\to 1^-} ={N^2 \ov 8}\ ,
\\
& {\partial^3 \over \partial t^3}\ln {d_n [\rho (k)]_{\rm max}\over n!}\Big|_{t\to 1^+} = {N^2 \ov 16}\ .
\end{split}
\ee
Therefore, this is a 3rd-order phase transition.

The picture that emerges in terms of fermion momenta is that, for large $N$, the singlet irrep corresponds
to a filled Fermi sea with
Fermi level $k=N$. Multiplying with fundamental irreps excites the state by one unit of momentum per
irrep and results in excitations around the Fermi level.  As long as $n<N^2/4$ the excitations remain
localized around the Fermi level and are $N$-independent. For $n=N^2/4$ the excitations reach the
bottom of the sea ($k=0$), marking a phase transition, and for $n>N^2/4$ the entire Fermi sea is excited
and lifted above $k=0$.
We remark that, in the case $1\ll 2\sqrt{n} <N <n$, there are in principle irreps with all the $k_i$ excited
above their ground state (singlet) values, but such irreps have subleading multiplicities and are irrelevant
in the large-$N$ limit.

\subsection{General repulsion $w\neq 1$, maximal $S_{w,n}$}

For general $w$, the equation for $\rho(k)$ that maximizes $S_{w,n} [\rho(k)]$ is \eqn{eomr}
with the constraints (\ref{constr}).
The solution proceeds along similar lines as the $w=1$ case. Rather than "duality 
preserving" and "duality violating" phases, we will talk about "condensed" and "dilute" cases, the
former being one where the density reaches its saturation value $\rho(k) =1$ for a range of values
of $k$, the latter one with $\rho(k)$ always less than 1. As we shall see, for $w>1$ the condensed phase
always involves a saturation region $[0,a]$ for $k$, while for $w<1$ it can saturate in a region
$[a_1 , a_2]$ with $0<a_1 <a_2$. We shall focus on the physically more relevant case $w\geqslant 1$
from now on, for reasons explained below \eqn{mww}.

\subsubsection{Dilute phase $n>n_w$}

In this case $\rho(k) <1$ for all $k$ and the solution can be obtained from the $w=1$ solution
with a simple rescaling. Specifically, 
\be
{\bar \rho}(k) = w \rho(k) \ ,\quad {\overline N} = w N \ ,\quad {\bar n} = wn +{w(1-w)N^2\over 2}\ ,
\ee
satisfy the same equations and constraints for $w=1$. We can use the solution in that case
to find
\be
\rho(k) = {2\over w\pi} \cos^{-1} {\sqrt{k}+\sqrt{ab/k} \over \sqrt{a}+\sqrt{b}}\ ,
\qq a\leqslant k\leqslant b\ ,
\label{wrhocos}
\ee
with
\be
\label{ablaw}
\begin{split}
&
\sqrt{a}=\sqrt{{(2-w)N\over 4}+{n\over N}}-\sqrt{wN\over 2} \ ,
\\
& \sqrt{b} = \sqrt{{(2-w)N\over 4}+{n\over N}}+\sqrt{wN\over 2}  \ ,
\\
&
\lambda = -\ln \left({(2-w)N\over 4} +{n \over N}\right) \ .
\end{split}
\ee
Clearly $\r (k) < w^{-1}$ and thus $\r (k) \leqslant 1$ for $w\geqslant  1$. The above solution will exist as long as
$\sqrt a \geqslant 0$, and thus for $n$ above a critical value $n_w$
\be
\label{nnw}
n > n_w = {(3w-2)N^2 \over 4}\quad \Longrightarrow \quad t> 3 w -2\ ,
\ee
for the order parameter $t$ defined in \eqn{torde}.
For $w=1$ we recover the transition at $n=N^2/4$, while for the number of states case $w=2$
the transition happens at $n=N^2$.

\subsubsection{Condensed phase $n<n_w$}

For $n<n_w$ ($t<3 w-2$) the solution (\ref{wrhocos}) ceases to exist and we enter the condensation phase. Similar to the solution \eqn{fjhq2}, we set
\be
\rho(k) = \left\{ \begin{array}{cc}
1\ , \qquad &0<k<a\ , \cr
\r_0 (k-a)\ , \qquad & a<k<a+b \ ,
\end{array} \right.
\ee
and zero elsewhere, with $a,b$ two positive constants. The function $\r_0 (k)$ satisfies the constraints
\be
0\leqslant  \rho_0 (k) \leqslant 1 \ ,\quad \int_0^b dk\, \rho_0 (k) = N-a\ ,\quad  
\int_0^b dk \, k\, \rho_0 (k) = {(N-a)^2 \over 2} +n\ .
\label{cono}
\ee
Substitution of $\rho (k)$ in (\ref{Sw}) yields, upon changing variable $k \to k+a$,
\be
\begin{split}
& S_{w,n} [\rho_0 (k)] = {w\ov 2} \int_0^\infty \bb\bb dk \int_0^\infty \bb dk' \, \rho_0 (k) \rho_0 (k') \ln |k - k' |
\\
&\quad -w \int_0^{\infty} \bb dk\, \rho_0(k)\, k (\ln k -1) + (w-1) \int_0^{ \infty} dk\, \rho_0 (k)\, (k+a)  \big(\ln (k+a) -1\big) \ ,
\end{split}
\ee
where we omitted terms set to be constant by the constraints and harmlessly extended the integration range
to infinity since $\rho_0 (k)$ has finite range.
Adding to this the constraints \eqn{cono} with appropriate Lagrange multipliers and following the procure that led to \eqn{r2} we 
obtain the equilibrium equation for $\r_0$
\be
w \int_0^\infty\! dk'\, \rho_0 (k') \ln|k-k'| = w k(\ln k -1)  -(w-1) (k+a)\big(\ln (k+a) -1\big)  
+ \mu +\lambda k\ .
\label{r20}
\ee
Taking the $k$-derivative we obtain the analog of \eqn{eomr}, i.e. 
\be
\int_0^\infty dk' {\rho_0 (k')\over k - k'} = \ln k -\gamma \ln(k+a) + \lambda/w\ , \qq{\rm when}
\ \  \rho_0 (k) > 0\ 
\label{wo}
\ee
and  where we defined for convenience the parameter\footnote{A similar equation to \eqn{wo} was found in 
 \cite{Betzios:2022pji} (eq. (5.2) of that work) in a matrix model approach to black  hole microstates.}
\be
\gamma = 1-{1\over w}\ ,\quad \g \in [0,1]\ .
\ee
We see that the equation for $\r_0(k)$ now has a two-logarithm potential, given by the right hand side of
\eqn{r20} as
\be
V_{\l,w,a}(k) =  w k(\ln k -1)  -(w-1) (k+a)\big(\ln (k+a) -1\big)   +\lambda k\ .
\ee
(up to a constant term $\mu$), while for $w=1$ the second logarithm drops out.

\no
To solve for $\r_0 (k)$, we define as before the resolvent
\be
u_0 (z) = \int dk\, {\rho_0 (k) \over z-k}\ ,
\label{uoz}
\ee
reproducing $\rho_0 (k)$ and its Hilbert transform as
\be
u_0 (k+i\epsilon) =  \Xint{-} dk' {\rho_0 (k') \over k-k'} -i\pi  \rho_0(k)\ .
\label{ukw}
\ee
In analogy to \eqn{rescut}, we set
\be
u_0 (z) = {1\over 2\pi i} \sqrt{z(z-b)} \oint ds\, {\ln s - \gamma \ln(s+a) 
+ \lambda/w \over (s-z) \sqrt{s(s-b)}}\ ,
\label{resw}
\ee
where the contour winds in the clockwise direction around the cut of the square root $[0,b]$
but does not include the
singularity at $z$ nor the cuts of the logarithms (so it "threads" the real line at $k=0$) (see fig. \ref{contour2}).
\begin{center}
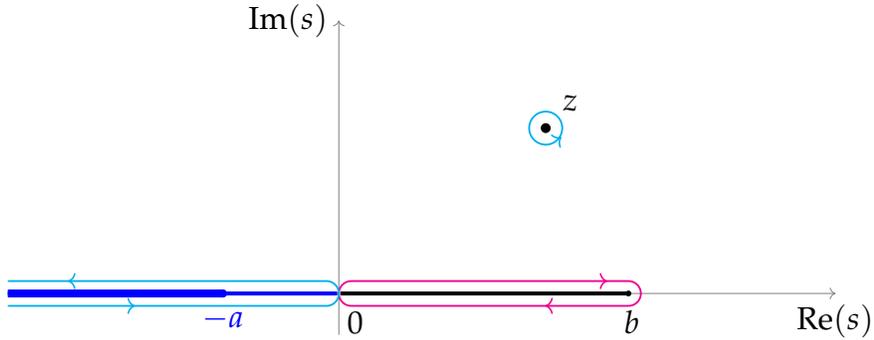

\begin{tikzpicture}[scale=1.1, decoration={markings,
mark=at position 0.9cm with {\arrow[line width=0.6pt]{<}},
mark=at position 7.4cm with {\arrow[line width=0.6pt]{<}}
}
]
\draw[help lines,->] (-4,0) -- (6,0) coordinate (xaxis);
\draw[help lines,->] (0,-0.5) -- (0,3.3) coordinate (yaxis);

\fill (2.5,2.0) circle (1.7pt);\fill (3.5,0) circle (1.0pt);\fill (-1.4,-0.0) [blue] circle (1.4pt);

\node at (0.2,-0.35) {$0$};
\node at (2.8,2.3) {$z$};
\node at (3.54,-0.34) {$b$};
\path[draw,line width=0.7pt,postaction=decorate] (2.3,2.0) [cyan] arc(180:-180:0.2);

\path[draw,line width=1.50pt] (0.02,0) 
-- (3.5,0)
;

\path[draw,line width=0.7pt,postaction=decorate] (3.5,-0.15) [magenta] arc(-90:90:0.15)
-- (0.15,0.15) arc(90:270:0.15) -- (3.5,-0.15) ;

\path[draw,line width=1.5pt] [blue] (-4,0.00) -- (-0.02,0);
\path[draw,line width=0.7pt,postaction=decorate] (-4,0.15) [cyan] -- (-0.15,0.15) arc(90:-90:0.15) -- (-0.15,-0.15) -- (-4,-0.15);

\path[draw,line width=3pt] (-4,-0.0) [blue] -- (-1.4,-0.0) node[below] {$-a$};

\node[below] at (xaxis) {$\text{Re}(s)$};
\node[left] at (yaxis) {$\text{Im}(s)$};

\end{tikzpicture}
\vskip -.4 cm
\captionof{figure}{\small{Contour of integration in the $s$ plane. The original (magenta) contour
around the (black) square root cut on $(0,b)$ is pulled back to the two (cyan) contours around the pole at $z$
and the two (blue) logarithm cuts on $(-\infty,0)$ and $(-\infty,-a)$.}
\label{contour2}}
\end{center}
Pulling back the contour we pick up the pole at $s=z$
and the integral around the cuts of the logarithms and obtain
\be
u_0  (z) = \ln z -\gamma \ln (z+a) +{\lambda\ov w} -\sqrt{z(z-b)}\left[\int_0^\infty - \gamma \int_a^\infty
\right] {ds \over (s+z)\sqrt{s(s+b)}} \ .
\label{resw1}
\ee
For $z =k$ real and between $0$ and $b$ (the region in which $\rho_0(k)$ does not vanish) the last
term proportional to $\sqrt{z(z-b)}$ is purely imaginary and, according to (\ref{ukw}), determines
$\rho_0(k)$ as
\be
\r_0 (k) = \sqrt{k(b-k)}\left[(1-\g)\int_0^\infty + \gamma \int_0^a \right] {ds \over (s+k)\sqrt{s(s+b)}} \ .
\ee
Upon performing the integrals,
\be
\r_0 (k) = {2\over w\pi} \cos^{-1} \sqrt{k \over b} + {2(w-1)\over w \pi}
\cos^{-1}\sqrt{(a+b) k \over (a+k)b}\ .
\ee
The density $\r_0(k)$ is a decreasing function of $k$, with $\rho_0 (0) =1$ and $\rho_0 (b) = 0$.
For $w=1$ this recovers the result (\ref{rcrit}), while for the number of states case $w=2$ it reduces to
\be
\r_0 (k) = {1\over \pi} \cos^{-1} \sqrt{k \over b} + {1\over \pi}
\cos^{-1}\sqrt{(a+b) k \over (a+k)b} = {1\over \pi} \cos^{-1} {(1 + \sqrt{1+b/a})k - b
\over b\sqrt{1+k/a}}\ .
\ee
The parameters $a$, $b$ and $\lambda$ can be related to $N$ and $n$ by matching the asymptotic 
expansion of $u_0 (z)$ 
\be
\label{assym2}
\begin{split}
u_0(z) &=  z^{-1} \int_0^\infty dk\, \rho_0(k) + z^{-2} \int_0^\infty dk\, k\, \rho_0(k) + {\cal O}\big(z^{-3}\big)\, 
 \\
&= z^{-1}\, (N-a) + z^{-2}\, \bigg(n+{(N-a)^2\ov 2}\bigg) +{\cal O}\big(z^{-3}\big)\ , 
\end{split}
\ee
from (\ref{uoz}) using (\ref{cono}) to those from (\ref{resw}). We obtain
\be
\label{laaa}
\begin{split}
& \lambda =  2 (w-1)\ln{\sqrt a + \sqrt{a+b} \over 2} -w \ln {b\ov 4} \ ,
\\
& N-a = {b\over 2} - {\gamma \over 2} \left(\sqrt{a+b}-\sqrt a\right)^2\ ,
 \\
& {(N-a)^2 \over 2} + n ={3b^2\over 16} + {\gamma \over 4} \left(-{3b^2\over 4}+2 a^2
+(b-2a)\sqrt{a(a+b)}\right)\ .
\end{split}
\ee
Although the system \eqn{laaa} is quite complicated, it can be explicitly solved for $a,b$. We set 
\be
\begin{split}
& a = N\, \biggl(\sqrt{\Big({w\ov 2}-1\Big)x +1}-\sqrt{{w x\ov 2}}\biggr)^2\ ,
\\
& b = 2N \sqrt{w x\Bigl((w-2)x+2\Bigr)}\ .
\end{split}
\ee
These satisfy the last two of \eqn{laaa} provided that
\be
x ={4 n/N^2\ov w+\sqrt{w^2+4(w-2)n/N^2}}
\ .
\ee
The above give for $\lambda$
%
\be
\label{llfin}
\begin{split}
\l=& -\ln {N\ov 2} -{w\ov 2}\ln \Big({4wn\ov N^2}\Big) +\Big({w\ov 2}-1\Big)  \ln \left(2-w+
\sqrt{w^2+(w-2){4n/ N^2}}\right)
\\
& \qq\qq\qq +{w\ov 2} \ln \left(w+\sqrt{w^2+(w-2){4n/N^2}}\right)\ .
\end{split}
\ee
Note that the argument of the square roots are strictly positive for $n<n_w$.

\no
As a consistency check, for $w=1$ the above reproduce the results (\ref{Nnla}) (for $a=0$),
while for $n=n_w =(3w-2)/4 N^2$, $x=1$ and thus $a=0$, $b=2wN$ and the results match
the results of the dense case at the critical point.
Also, $n=0$ implies $x=0$, and thus $a=N$, $b=0$, reproducing the singlet distribution.

\no
For $w=2$ the results simplify considerably and we obtain
\be
\label{abl}
w=2:\qq a = \biggl(\sqrt {N} - \sqrt {n\ov N}\biggr)^2
\ ,\qquad  b = 4\sqrt {n} \ , \qquad \l = -\ln {n\ov N}\ .
\ee
Remarkably, these are just the dilute case results \eqn{ablaw} for $w=2$ (noting that $b$ in that
case maps to $a+b$ in the present case), analytically continued for negative values of $n-N^2$.

\begin{figure} [th!]
\begin{center}
\hskip -0.6cm\includegraphics[height= 7 cm, angle=0]{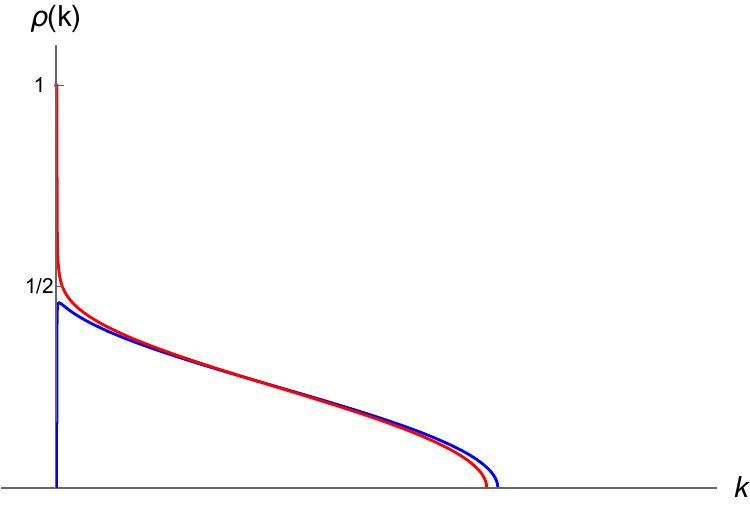}
\end{center}
\vskip -.5 cm
\caption{\small{The distribution $\rho(k)$ for $w=2$ around the critical point $n=N^2$. Both the dilute
(blue) density and the dense (red) density have a value close to $w^{-1} = 1/2$ near $k=0$, but the dilute one
sharply dips to $0$ and the dense one sharply rises to $1$, for an increasingly sharp transition as $n$ crosses
the critical value $N^2$.}}
\label{Rhw}
\end{figure}

The behavior of $\rho (k)$ for $w>1$ is qualitatively similar as for $w=1$, with the notable exception
that the maximal value of $\rho(k)$ in the dilute phase is $\rho = 1/w$, achieved for $k \simeq 0$ as
$n \to n_w +\epsilon$. However, $\r (0)$ jumps from $\rho(0) = 0$ for $n=n_w +\e$ to $\rho(0) = 1$
for $n=n_w-\e$. This behavior is displayed in fig. \ref{Rhw}.

\subsubsection{Phase transition} 

The general $w$ case exhibits an interesting pattern of phase transitions, depending on the value of $w$.
As in the $w=1$ case, the "free energy" functional $S_w [\r(k)]$ appearing in \eqn{Sw} does not depend
explicitly on $n$, so its derivative with respect to $n$ at the dominant configuration is still given by the value of the
Lagrange multiplier $\lambda$. We have explicit expressions for $\lambda$ in both phases, dilute
\eqn{ablaw} and condensed \eqn{llfin}, so we may directly calculate its derivatives on either side
of the critical point $n=n_w$. Equivalently, we give its expansion around the critical point. For the
condensed phase $n<n_w$ the expansion is
\be
\label{lsxp1}
\begin{split}
n<n_w :\qq & \l = -\ln{w N\ov 2}  - {2\ov w N^2} (n-n_w) + {2\ov w^2 N^4}\, (n-n_w)^2 
\\
& \qq\qq\qq -  {8- 14 w + 7 w^2 \ov 3 w^3 (1-w)^2 N^6} \, (n-n_w)^3+\dots  
\end{split}
\ee
while for the dilute phase $n>n_w$ we obtain 
\be
\label{lsxp2}
\begin{split}
n>n_w :\qq& \l = -\ln{w N\ov 2}  - {2\ov w N^2} (n-n_w) + {2\ov w^2 N^4}\, (n-n_w)^2  
\\
& \qq\qq\qq\qq -  {8 \ov 3 w^3  N^6} \, (n-n_w)^3+\dots  \ .
\end{split}
\ee
We see that for all values of $w>1$, except $w=2$, $\lambda$ and its first two derivatives in $n$
(equivalently in $t$) are continuous,
while its third derivative is discontinuous at $n=n_w$. This means that the first three derivatives of $S_w$
are continuous but the {\it fourth} one is discontinuous, signaling a {\it fourth-order} phase transition
at $n=n_w$.\footnote{In \cite{Betzios:2022pji} a third order phase transition was mentioned with an order parameter related to $w$. Its relationship to our present work is not obvious.}

The values $w=2$ and $w=1$ are special. For $w=2$, {\it all} derivatives of $\lambda$ are continuous
across $n=n_2 = N^2$, since $\lambda$ in this case is given by a unique analytic function of $n$
\eqn{abl}, so there is {\it no} phase transition.
\begin{figure} [th!]
\begin{center}
\hskip -0.6cm\includegraphics[height= 7 cm, angle=0]{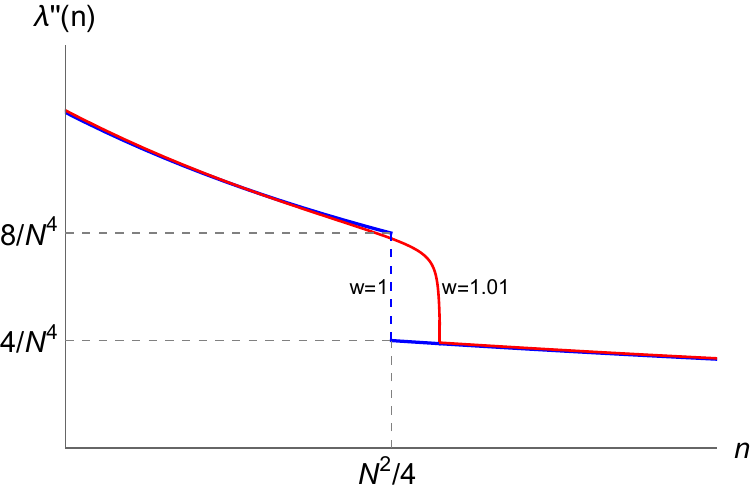}
\end{center}
\vskip -.5 cm
\caption{\small{Plots of $\lambda'' (n) =d^2\lambda/dn^2$ for $w=1$ (blue curve) and $w=1.01$ (red curve)
around the $w=1$ critical point $n=N^2/4$. For $w=1.01$ there is a sharp transition from 
$\lambda'' \simeq 4/N^4$ to $\lambda'' \simeq 8/N^4$ but no discontinuity, while for $w=1$ the transition
evolves into a discontinuity.
The $w=1.01$ curve has a cusp at its critical point, signifying a discontinuous 3$^{rd}$ derivative and a
fourth-order phase transition.}}
\label{w1}
\end{figure}

The case $w=1$ is trickier. From \eqn{lsxp1} and \eqn{lsxp2} we would deduce that the second
derivative of $\l$ is continuous, in contradiction to the results of section {\bf \ref{transition1}}.
What in fact happens is that for $w=1+\epsilon$, the second derivative $d^2\l / dn^2$ in the
condensed phase displays an increasingly sharp transition: it approaches $8/N^4$ as $n$ approaches
$n_1 = N^2/4$, which is the value for $w=1$, but at $n\sim N^2/4 -\epsilon^2$ it starts sharply dropping,
and as $n$ exceeds $N^2/4$ it reaches $4/N^4$ , the value consistent with \eqn{lsxp1} (see fig. \ref{w1}).
Physically, for $w$ close to $1$ the system goes through a crossover near $n=n_w$,
which is increasingly sharp as $w$ approaches 1 and becomes a full phase transition at $w=1$.
Mathematically, the limits $w\to 1$ and $n \to n_w$ in the
condensed phase do not commute. By contrast, in the dilute phase the $w\to 1$ limit is smooth.
The various phase transitions are summarized in table \ref{table:1}.
\begin{table}[ht]
\begin{center}
\begin{tabular}{|c|c|c|c|} \hline
  transition & 3rd order&  4th order & no transition
  \\ \hline \hline
$w=1$  &  $\checkmark$    &     &  
\\ \hline
$w>1\, (w\neq 2)$ & \rm{(crossover)}  &  $\checkmark$    &   
 \\ \hline
 $w=2$ &   &   & $\checkmark$    
  \\\hline
\end{tabular}
\end{center}
\vskip -.3 cm
\caption{ \small{Phase transitions for various values of $w\geqslant 1$.}}
\label{table:1}
\end{table}
\no
As in the $w=1$ case, we can integrate $\lambda$ with respect to $n$ to find the quantity $m_{w,n}$
for general $w$. Note that the relation of $m_{w,n}$ of \eqn{mww} with $S_{w,n}$ of \eqn{Sw} is 
\be
\ln m_{w,n} = S_{w,n} + \ln n! -(w-1) \sum_{s=1}^{N-1} \ln s! 
\ee 
and is defined such that for the singlet irrep it becomes $1$. This fixes the integration constant in the
dense case, and continuity at $n=n_w$ fixes the integration constant in the dilute case. We obtain
for the dense phase $n<n_w$
\be
\begin{split}
&\ln m_{w,n;\hskip 0.03cm\rm{max}}= \frac{2 n (w-1)}{w+\sqrt{\frac{4 n (w-2)}{N^2}+w^2}}
-{n\over 2}\bigl( (w-2)\ln n +w\ln w +1 \bigr)
\\
&\qq  +\ha \left(w \left(n+N^2\right)-2 n-N^2\right) \ln \biggl(\sqrt{\frac{4 n (w-2)}{N^2}+w^2}-w+2\biggr)
\\ 
&\qq +{n w\ov 2} \ln \biggl(\sqrt{\frac{4 n (w-2)}{N^2}+w^2}+w\biggr) +
(w-1) \biggl(n\ln{N\over 2} - {N^2\over 2} \ln 2 \biggr)\ ,
\end{split}
\label{mden}
\ee
which indeed vanishes for $n=0$. For the dilute phase $n > n_w$ we have that
\be
\begin{split}
&\ln m_{w,n;\hskip 0.03cm\rm{max}} = n\ln n - N  \left({(2-w)N\over 4} +{n \over N}\right) \ln \left({(2-w)N\over 4} +{n \over N}\right)
\\
& \qq\qq\qq
+{N^2 \ov 8}\Big(4 \ln N + w-2 + 2 w \ln {w\ov 2 N}\Big)\ .
\end{split}
\label{mdil}\ee
In the special case $w=2$ the expressions \eqn{mden} and \eqn{mdil} simplify and unify, giving the result
\be
m_{2,n;\hskip 0.03cm\rm{max}} = {N^n}\ ,
\ee
as expected, since the saddle point in the large-$N,n$ limit must give the full number of states $N^n$ up
to subleading terms (determinants).

We conclude with a couple of remarks. In the case $0<w<1$, $\rho (k)$ in
\eqn{wrhocos} can reach or exceed the value 1 for $n$ below a critical value
{\it higher} than $n_w$, or equivalently a critical value for $t$
\be
t < t_c = {2w \over \sin^{2}{\pi w\over 2}} + w -2 \ .
\ee
For $t< t_c$, or $n <t_c N^2 /4$, the solution \eqn{wrhocos} is not valid any more, as it exceeds
1 at some interval. The true solution is one with
\be
\rho(k) =  \left\{ \begin{array}{cc}
0\ , ~~ & ~~~ 0<k<a_1 ~~\text{or}~~ a_2<k \ ,
\cr
\rho_1 (k)\ ,  ~~ &~~~ a_1<k<b_1 \ ,
\cr
1\ ,  ~~ &~~~ b_1<k<b_2 \ ,
\cr
\rho_2 (k)\ ,  ~~ &~~~ b_2<k<a_2 \ .
\end{array} \right.
\ee
leading to a genuine two-cut Cauchy problem. We will not explore this solution, as $w<1$ would correspond
to a statistical model with entropy decreasing as the dimensionality of the irrep increases (see \eqn{mww}).
This is rather unphysical, although it could conceivably find application in some exotic situation.

Finally, the case $w<0$ is even more unphysical, representing a drastic reduction of entropy.
Its large-($N,n$) limit would correspond to $N$ particles {\it attracting} each other with two-body
logarithmic potentials. In the concave external potential $V_\lambda (k)$ of \eqn{potvl} the only stable
configuration is one with all particles coalescing to an
interval of length $N$, for a density of $1$, corresponding to the singlet irrep.

\section{Conclusions}

We considered the multiplicity of irreps arising in the decomposition of $n$ fundamental
representations of $SU(N)$, weighted by a power of their dimension.
We showed that a nontrivial double scaling limit exists in which both $n$ and $N$ become large keeping
the ratio $n/N^2$ fixed, and uncovered novel phase transitions in which this ratio plays the role of the order
parameter. The system generically undergoes a fourth order phase transition, from
a dense to a dilute phase, enhanced to a third order one for the unweighted multiplicity, and ceasing to exist
altogether when weighting with the first power of the dimension, which corresponds to the infinite
temperature partition function of nonabelian ferromagnets.  

Our results are model independent, not involving a Hamiltonian, and should thus be relevant to the
thermodynamics of nonabelian ferromagnets at high temperatures. In this respect, it is interesting to
reconsider the phase structure of the ferromagnetic model in which $n$ atoms mutually interact via
$SU(N)$ components, which we recently investigated in \cite{Phases} in the thermodynamic limit $n \gg 1$
but for fixed finite $N$. We expect that the present double scaling limit will qualitatively modify the phase
structure, provided $N \sim \sqrt n$. The generalization of our results for a product of irreps other than
the fundamental would also be interesting, the adjoint being the most natural alternative choice. The
related question of the relation of our results to Markov processes in the space of tableaux, such as
the Plancherel process, is also of mathematical interest.

Finally, the relevance of our results to matrix models and large-$N$ Yang-Mills theories should be explored.
Of particular interest is the understanding of microstates in the two-dimensional black hole 
of \cite{Witten:1991yr, Mandal:1991tz} with matrix models, along the lines of 
\cite{Kazakov:2000pm,Kazakov:2001pj} and more recently of \cite{Betzios:2022pji, Ahmadain:2022gfw},
and of the deconfinement/Hagedorn transition in large-$N$ gauge theories \cite{Bo,AMMPR,HaMaSu}, especially
in the setting of \cite{Hanada,ark}. These and other related questions are the subject of ongoing investigation.

\subsection*{Acknowledgements}

We would like to the thank David Berenstein for a useful correspondence.\\
The research of A.P. was supported by the National Science Foundation 
under grant NSF-PHY-2112729 and  by PSC-CUNY grants 65109-00 53 and 6D136-00 02.\\
The research of K.S. was supported by the Hellenic Foundation for
Research and Innovation (H.F.R.I.) under the ``First Call for H.F.R.I.
Research Projects to support Faculty members and Researchers and
the procurement of high-cost research equipment grant'' (MIS 1857, Project Number: 16519).



\begin{thebibliography}{99.}

\bibitem{Hoo}
G.~'t Hooft, {\it A Planar Diagram Theory for Strong Interactions}, \hfill\break 
\href{https://doi.org/10.1016/0550-3213(74)90154-0}{Nucl. Phys. \textbf{B72} (1974) 461}.

\bibitem{BIPZ}
E.~Brezin, C.~Itzykson, G.~Parisi, and J.B.~Zuber, {\it Planar Diagrams}, \hfill\break 
\href{https://link.springer.com/article/10.1007/bf01614153}{Commun. Math. Phys. \textbf{59} (1978) 35}.

\bibitem{GrMi}
D.J.~Gross and A.A.~Migdal, {\it Nonperturbative two-dimensional quantum gravity},
\href{https://journals.aps.org/prl/abstract/10.1103/PhysRevLett.64.127}{Phys. Rev. Lett. \textbf{64} (1990) 127}.\hfill\break
M.R.~Douglas and S.H.~Shenker, {\it Strings in Less Than One Dimension},\hfill\break
\href{https://doi.org/10.1016/0550-3213(90)90522-F}{Nucl. Phys. \textbf{B335} (1990) 635}.\hfill\break
I.I. Lungu, A.M. Grumezescu and C. Fleaca, {\it Unexpected Ferromagnetism $-$ A Review,}
\href{https://doi.org/10.3390/app11156707}{Appl. Sci. 2021, 11(15), 6707}.

\bibitem{Ghu}
M.A. Cazalilla, A.F. Ho and M. Ueda,
{\it Ultracold gases of ytterbium: ferromagnetism and Mott states in an $SU(6)$ Fermi system},
\href{https://iopscience.iop.org/article/10.1088/1367-2630/11/10/103033}{2009 New J. Phys. {\bf 11} 103033}.

\bibitem{Gor}
A.V.~Gorshkov {\it et al.}, {\it Two-orbital SU(N) magnetism with ultracold alkaline-earth atoms},
\href{https://www.nature.com/articles/nphys1535}{Nature Physics \textbf{6} (2010) 289-295}.

\bibitem{Zha}
X. Zhang  {\it et al.}, {\it Spectroscopic observation of $SU(N)$-symmetric interactions in Sr orbital
magnetism}, \href{https://www.science.org/doi/abs/10.1126/science.1254978}
{Science Vol 345, Issue 6203 (2014) 1467}.

\bibitem{Mag}
M. A. Cazalilla and A.M. Rey, {\it Ultracold Fermi gases with emergent $SU(N)$ symmetry}, 
\href{https://iopscience.iop.org/article/10.1088/0034-4885/77/12/124401}{2014 Rep. Prog. Phys. {\bf 77} 124401}.

\bibitem{Cap}
S.~Capponi, P.~Lecheminant, and K.~Totsuka, {\it Phases of one-dimensional SU(N) cold atomic 
Fermi gases - From molecular Luttinger liquids to topological phases}, \hfill\break 
\href{https://www.sciencedirect.com/science/article/abs/pii/S0003491616000130}{Ann. Phys. \textbf{367} (2016) 50-95}.

\bibitem{Aff}
I.~Affleck, {\it Large-$n$ Limit of $SU(n)$ Quantum "Spin" Chains}, \hfill\break
\href{https://journals.aps.org/prl/abstract/10.1103/PhysRevLett.54.966}{Phys. Rev. Lett. \textbf{54} (1985) 966}.

\bibitem{Pola}
A.P.~Polychronakos, {\it Exact Spectrum of $SU(n)$ Spin Chain with Inverse-Square Exchange},
Nucl. Phys. \textbf{B419} (1994) 553-566, 
\href{https://arxiv.org/abs/hep-th/9310095}{arXiv:hep-th/9310095}.


\bibitem{KT}
H. Katsura and A. Tanaka,
{\it Nagaoka states in the $SU(n)$ Hubbard model}, \hfill\break
\href{https://journals.aps.org/pra/abstract/10.1103/PhysRevA.87.013617}{Phys. Rev. {\bf A87} (2013) 013617}.


\bibitem{BSL}
E. Bobrow, K. Stubis and Y. Li,
{\it Exact results on itinerant ferromagnetism and the 15-puzzle problem},
\href{ https://journals.aps.org/prb/abstract/10.1103/PhysRevB.98.180101}
{Phys. Rev. {\bf B98} (2018) 180101(R)}.

\bibitem{RoLa}
C. Romen and A.M. L\"auchli, {\it Structure of spin correlations in high-temperature $SU(N)$
quantum magnets},
 \href{https://journals.aps.org/prresearch/abstract/10.1103/PhysRevResearch.2.043009}{Phys. Rev. Research 2
(2020) 043009}.

\bibitem{YSMOF}
D. Yamamoto, C. Suzuki, G. Marmorini, S. Okazaki and N. Furukawa,
{\it Quantum and Thermal Phase Transitions of the Triangular $SU(3)$ Heisenberg Model under Magnetic Fields},
\href{https://journals.aps.org/prl/abstract/10.1103/PhysRevLett.125.057204} {Phys. Rev. Lett. {\bf 125} (2020) 057204}.

\bibitem{TK}
 K. Tamura and H. Katsura, 
{\it   Ferromagnetism in $d$-Dimensional $SU(n)$ Hubbard Models with Nearly Flat Bands},
\href{https://link.springer.com/article/10.1007/s10955-020-02687-w}{Journal of Stat. Phys. volume {\bf 182}, 16 (2021)}.


\bibitem{Totsuka} K. Totsuka, 
{\it Ferromagnetism in the $SU(N)$ Kondo lattice model: $SU(N)$ double exchange and supersymmetry}
\href{https://journals.aps.org/pra/abstract/10.1103/PhysRevA.107.033317} {Phys. Rev. {\bf A107} (2023) 033317}.

\bibitem{TK2}
K. Tamura, H. Katsura,
{\it Flat-band ferromagnetism in the $SU(N)$ Hubbard and Kondo lattice models}
\href{https://arxiv.org/abs/2303.15820}{arXiv:2303.15820 [cond-mat]}.

\bibitem{DY}
D.~Yamamoto {\it et al.}, {\it Quantum and Thermal Phase Transitions of the Triangular $SU(3)$ Heisenberg Model under Magnetic Fields}, Phys. Rev. Lett. \textbf{125} (2020) 057204.

\bibitem{YM}
Y.~Miyazaki {\it et al.},
{\it Linear Flavor-Wave Analysis of $SU(4)$-Symmetric Tetramer Model with Population Imbalance},
J. Phys. Soc. Jpn. {\bf 91} (2022) 073702.

\bibitem{HM}
H.~Motegi {\it et al.}, {\it Thermal Ising transition in
two-dimensional $SU(3)$ Fermi lattice gases with population imbalance}, 
\href{https://arxiv.org/abs/arXiv:2209.05919}{arXiv:2209.05919 [cond-mat]}.

\bibitem{Polychronakos:2023yhq}
A.~P.~Polychronakos and K.~Sfetsos,
{\it Composing arbitrarily many $SU(N)$ fundamentals},
Nucl. Phys.  \textbf{B994} (2023) 116314,
\href{https://arxiv.org/abs/arXiv:2305.19345}{arXiv:2305.19345 [hep-th]}.

\bibitem{Phases}
A.P.~Polychronakos and K.~Sfetsos, {\it Ferromagnetic phase transitions in $SU(N)$},
Nucl. Phys.  \textbf{B996} (2023) 116353,
\href{https://arxiv.org/abs/arXiv:2306.01051}{arXiv:2306.01051 [hep-th]}.

\bibitem{MiPo}
J.A.~Minahan and A.P.~Polychronakos, {\it Equivalence of Two Dimensional QCD and the $c=1$ Matrix Model},
Phys. Lett. \textbf{B312} (1993) 155-165,
\href{https://doi.org/10.48550/arXiv.hep-th/9303153}{arXiv:hep-th/9303153}.

\bibitem{Polb}
For a pedagogical review see A.P.~Polychronakos, {\it Physics and Mathematics of Calogero particles}, 
J. Phys. \textbf{A39} (2006) 12793-12846, \href{https://arxiv.org/abs/hep-th/0607033}{arXiv:hep-th/0607033}.

\bibitem{VK}
A.M.~Vershik and S.V.~Kerov, {\it Asymptotics of the Plancherel measure of the symmetric group and the
limiting form of Young tableaux}, \hfill\break
\href{https://www.mathnet.ru/php/archive.phtml?wshow=paper&jrnid=dan&paperid=40430&option_lang=eng}{Soviet Mathematics Doklady \textbf{18} (1977) 527-531}.

\bibitem{LS}
B.F.~Logan and L.A.~Shepp, {\it A variational problem for random young tableaux}, \hfill\break
\href{https://www.sciencedirect.com/science/article/pii/0001870877900305}{Adv. Math. \textbf{26} (1977) 206-222}.

\bibitem{Betzios:2022pji}
P.~Betzios and O.~Papadoulaki, {\it Microstates of a 2d Black Hole in string theory},\hfill\break
JHEP \textbf{01} (2023) 028,
\href{https://arxiv.org/abs/arXiv:2210.11484}{arXiv:2210.11484 [hep-th]}.

\bibitem{Witten:1991yr}
E.~Witten,
{\it On string theory and black holes}, 
\href{https://journals.aps.org/prd/abstract/10.1103/PhysRevD.44.314} {Phys. Rev. \textbf{D44} (1991), 314-324}.

\bibitem{Mandal:1991tz}
G.~Mandal, A.~M.~Sengupta and S.~R.~Wadia,
{\it Classical solutions of two-dimensional string theory},
\href{https://www.worldscientific.com/doi/abs/10.1142/S0217732391001822 }{Mod. Phys. Lett. \textbf{A6} (1991), 1685-1692}.

\bibitem{Kazakov:2000pm}
V.~Kazakov, I.~K.~Kostov and D.~Kutasov,
{\it A Matrix model for the two-dimensional black hole},
Nucl. Phys. \textbf{B622} (2002), 141-188,
\href{https://arxiv.org/abs/hep-th/0101011}{arXiv:hep-th/0101011}.

\bibitem{Kazakov:2001pj}
V.~A.~Kazakov and A.~A.~Tseytlin,
{\it On free energy of 2-D black hole in bosonic string theory},
JHEP \textbf{06} (2001), 021,
\href{https://arxiv.org/abs/hep-th/0104138}{arXiv:hep-th/0104138}.

\bibitem{Ahmadain:2022gfw}
A.~Ahmadain, A.~Frenkel, K.~Ray and R.~M.~Soni,
{\it Boundary Description of Microstates of the Two-Dimensional Black Hole},
\href{https://arxiv.org/abs/arXiv:2210.11493}{arXiv:2210.11493 [hep-th]}.

\bibitem{Bo}
B.~Sundborg, {\it The Hagedorn transition, deconfinement and $N = 4$ SYM theory},\hfill\break
Nucl. Phys. \textbf{B573} (2000) 349,
\href{https://arxiv.org/abs/hep-th/9908001}{arXiv:hep-th/9908001}.


\bibitem{AMMPR}
O.~Aharony, J.~Marsano, S.~Minwalla, K.~Papadodimas, and M. van Raamsdonk, {\it The Hagedorn-deconfinement phase transition in weakly coupled large N gauge theories},
Adv. Theor. Math. Phys. \textbf{8} (2004) 603-696,
\href{https://arxiv.org/abs/hep-th/0310285}{arXiv:hep-th/0310285}.

\bibitem{HaMaSu}
M.~Hanada, J.~Maltz, and L.~Susskind, {\it Deconfinement transition as black hole formation by the condensation of
QCD strings}, Phys. Rev. D {\bf 90} (2014) 105019 

\bibitem{Hanada}
M.~Hanada, A.~Jevicki, C.~Peng, and N.~Wintergerst,
{\it Anatomy of deconfinement},
J. High Energ. Phys. {\bf 12} (2019) 167,
\href{https://arxiv.org/abs/arXiv:1909.09118}{arXiv:1909.09118 [hep-th]}.

\bibitem{ark}
D.~Berenstein and K.~Yan, {\it The endpoint of partial deconfinement},\hfill\break
\href{https://arxiv.org/abs/arXiv:2307.06122}{arXiv:2307.06122 [hep-th]}.






\end{thebibliography}
\end{document}